\documentclass[a4paper,11pt]{article}
\usepackage{jheppub}
\usepackage{amsmath,amssymb,latexsym}
\usepackage{braket}
\usepackage{bm}
\usepackage{booktabs}
\usepackage{setspace}
\usepackage{hepunits}
\usepackage[svgnames]{xcolor}
\usepackage{mciteplus}

\usepackage{tikz-cd}
\usetikzlibrary{patterns}
\usepackage{tikz}
\usetikzlibrary{arrows,shapes}
\usetikzlibrary{trees}
\usetikzlibrary{matrix,arrows} 				
\usetikzlibrary{positioning}				
\usetikzlibrary{calc,through}				
\usetikzlibrary{decorations.pathreplacing}  
\usepackage{pgffor}							

\usetikzlibrary{decorations.pathmorphing}	
\usetikzlibrary{decorations.markings}
\tikzset{
	photon/.style={decorate, decoration={snake}, draw=red},
	electron/.style={draw=blue, postaction={decorate},
		decoration={markings,mark=at position .55 with {\arrow[draw=blue]{>}}}},
	gluon/.style={decorate, draw=blue,
		decoration={coil,amplitude=4pt, segment length=4pt}} ,
	vector/.style={decorate, decoration={snake}, draw},
	provector/.style={decorate, decoration={snake,amplitude=2.5pt}, draw},
	antivector/.style={decorate, decoration={snake,amplitude=-2.5pt}, draw},
	fermion/.style={draw=black, postaction={decorate},
		decoration={markings,mark=at position .55 with {\arrow[draw=black]{>}}}},
	fermionbar/.style={draw=black, postaction={decorate},
		decoration={markings,mark=at position .55 with {\arrow[draw=black]{<}}}},
	fermionnoarrow/.style={draw=black},
	scalar/.style={draw=black, postaction={decorate},
		decoration={markings,mark=at position .55 with {\arrow[draw=black]{>}}}},
	scalarbar/.style={dashed,draw=black, postaction={decorate},
		decoration={markings,mark=at position .55 with {\arrow[draw=black]{<}}}},
	scalarnoarrow/.style={dashed,draw=black},
	electron/.style={draw=black, postaction={decorate},
		decoration={markings,mark=at position .55 with {\arrow[draw=black]{>}}}},
	bigvector/.style={decorate, decoration={snake,amplitude=4pt}, draw},
	background/.style={dashed,draw=black, postaction={decorate},
		decoration={markings,mark=at position 1 with {\arrow[draw=black]{<>}}}},
}

\tikzstyle{block} = [draw, rectangle, 
minimum height=3em, minimum width=6em]

\newcommand{\ep}{\varepsilon}
\newcommand{\nn}{\nonumber}
\newcommand{\nd}{\mathrm{d}}

\newcommand{\be}{\begin{equation}}
\newcommand{\ee}{\end{equation}}
\newcommand{\ba}{\begin{eqnarray}}
\newcommand{\ea}{\end{eqnarray}}

\allowdisplaybreaks

\def\eref#1{(\ref{#1})}
\newcommand{\bea}{\begin{eqnarray}}
	\newcommand{\eea}{\end{eqnarray}}
\newcommand{\bean}{\begin{eqnarray*}}
	\newcommand{\eean}{\end{eqnarray*}}

\def\ie{\begin{equation}\begin{aligned}}
\def\fe{\end{aligned}\end{equation}}

\def\bgm{\begin{matrix}}
\def\edm{\end{matrix}}
\def\bge{\begin{equation}}
\def\ede{\end{equation}}

\usepackage{framed}
\usepackage{mdframed}

\newmdenv[skipabove=0pt,%
          skipbelow=5pt,%
          leftmargin=0pt,%
          rightmargin=0pt,%
          innertopmargin=-5pt,%
          innerbottommargin=7pt,%
          innerleftmargin=2pt,%
          innerrightmargin=2pt,%
          splittopskip=0pt,%
          splitbottomskip=0pt,%
          linewidth=0pt,%
          nobreak=true]%
          {keyeqn2}

\newmdenv[backgroundcolor=gray!15,%
          skipabove=0pt,%
          skipbelow=5pt,%
          leftmargin=0pt,%
          rightmargin=0pt,%
          innertopmargin=-5pt,%
          innerbottommargin=7pt,%
          innerleftmargin=2pt,%
          innerrightmargin=2pt,%
          splittopskip=0pt,%
          splitbottomskip=0pt,%
          linewidth=0pt,%
          nobreak=true]%
          {keyeqn}

\title{
Multivariate hypergeometric solutions of cosmological (dS) correlators by $\text{d} \log$-form differential equations
}

\author[a,b,c]{Jiaqi Chen}
\author[c,d]{Bo Feng}
\author[e]{Yi-Xiao Tao}

\affiliation[a]{Beijing Key Laboratory of Optical Detection Technology for Oil and Gas, China University of Petroleum-Beijing, Beijing 102249, China}
\affiliation[b]{Basic Research Center for Energy Interdisciplinary, College of Science, China University of Petroleum-Beijing, Beijing 102249, China}
\affiliation[c]{Beijing Computational Science Research Center, Beijing 100084, China}
\affiliation[d]{Peng Huanwu Center for Fundamental Theory, Hefei, Anhui 230026, China}
\affiliation[e]{Department of Mathematical Sciences, Tsinghua University, Beijing 100084, China}

\emailAdd{jiaqichen@csrc.ac.cn}
\emailAdd{fengbo@csrc.ac.cn}
\emailAdd{taoyx21@mails.tsinghua.edu.cn}

\abstract{
In this paper, we give the analytic expression for the homogeneous part of solutions of arbitrary tree-level cosmological correlators, including massive propagators and time-derivative interaction cases. The solutions are given in the form of multivariate hypergeometric functions. It is achieved by two steps. Firstly, we indicate the factorization of the homogeneous part of solutions, i.e., the homogeneous part of solutions of multiple vertices is the product of the solutions of the single vertex. Secondly, we give the solution to the $\text{d} \log$-form differential equations of arbitrary single vertex integral family.  We also show how to determine the boundary conditions for the differential equations.
There are two techniques we developed for the computation. Firstly, we analytically solve $\text{d} \log$-form differential equations via power series expansion. Secondly, we handle degenerate multivariate poles in power series expansion of differential equations by blow-up. They could also be useful in the evaluation of multi-loop Feynman integrals in flat spacetime.
}

\begin{document}

\maketitle


\section{Introduction}
Anti-de Sitter (AdS) and de Sitter (dS) space as the simplest curved spacetime, quantum field theory (QFT) in them should be the first step for people to understand QFT in curved space. Meanwhile, QFT in dS space also catches people's interest due to its phenomenological application in cosmology, especially inflation physics. Inflation has been widely accepted as a period of the evolution of our early universe. During inflation, the background spacetime can be regarded as approximately a dS spacetime. The quantum fluctuation of all particles in the early universe gave rise to cosmic microwave background (CMB) and produced the Large Scale Structure (LSS) we can observe today. In order to obtain more information contained in the CMB and LSS, we need to analyze the cosmological correlators that are related to the Cosmological Collider signals \cite{Arkani-Hamed:2015bza,Chen:2009we,Chen:2009zp,Chen:2012ge,Chen:2016uwp}.

Cosmological correlators 
can be calculated by wavefunction coefficients or in-in Feynman rules. The former regards the cosmological correlator as inserting external fields in the field integral of the squared norm of the Hartle-Hawking wavefunction. The wavefunction coefficients, which encode all information of the wavefunction, are equivalent to the AdS amplitudes in the momentum space up to an analytic continuation. The latter is based on in-in formalism \cite{Keldysh:1964ud,Schwinger:1960qe,Chou:1984es,Penrose1986QuantumCI}. On the one side, many techniques analog to these methods in flat amplitudes and CFT correlators are developed in the calculation
, including cosmological bootstrap \cite{Arkani-Hamed:2018kmz,Baumann:2019oyu,Baumann:2020dch,Pajer:2020wnj,Hillman:2021bnk,Baumann:2021fxj,Hogervorst:2021uvp,Pimentel:2022fsc,Jazayeri:2022kjy,Wang:2022eop,Baumann:2022jpr,Chen:2023xlt} which involves some singularity behaviors and weight-shifting operators, off-shell methods \cite{Armstrong:2022mfr,Tao:2022nqc,Chen:2023bji,Werth:2024mjg}, family-tree decomposition \cite{Fan:2024iek} which could give power series solutions of arbitrary tree-level amplitude in conformal coupled case\footnote{Power series solutions also could be regarded as multivariate hypergeometric functions, and hypergeometric structure of Feynman integrals in flat cases and its evaluation also has been studied in \cite{Blumlein:2021hbq} recently.}, Mellin amplitudes \cite{Sleight:2019hfp,Sleight:2019mgd,Sleight:2020obc}, summation-by-parts relations in Mellin space \cite{Alaverdian:2024llo}, bootstrap equation \cite{Qin:2023ejc,Aoki:2024uyi},   (partial) Mellin-Barnes integration \cite{Sleight:2021plv,Jazayeri:2021fvk,Premkumar:2021mlz,Qin:2022lva,Qin:2022fbv,Qin:2023bjk}, spectral decomposition \cite{Xianyu:2022jwk,Liu:2024xyi,Loparco:2023rug} , Integrate-By-Part (IBP) \cite{Chetyrkin:1981qh} and the IBP-based differential equations \cite{Kotikov:1990kg,Kotikov:1991pm,Gehrmann:1999as,Bern:1993kr} for conformal coupled case \cite{De:2023xue,Arkani-Hamed:2023kig,He:2024olr,Benincasa:2024ptf} and general case \cite{Chen:2023iix} of dS background, and so on \cite{Mei:2023jkb,Gomez:2021qfd,Gomez:2021ujt,Arkani-Hamed:2017fdk,Arkani-Hamed:2018bjr,Lee:2022fgr,Grimm:2024tbg}. 

 This paper aims to show a systematic, powerful, and user-friendly method to evaluate perturbative QFT in dS space, thus cosmological correlators as well, and show the elegant structures of tree-level cosmological correlators. The method is mainly based on IBP and differential equations of general dS case \cite{Chen:2023iix}, including massive propagators and time derivative interaction.   This case is more non-trivial because people need to generalize IBP from the polynomial integrand case of flat or conformal coupled dS cases to the Hankel integrand case. Moreover, \cite{Chen:2023iix} directly gives the uniform formulas of iterative IBP reduction and $\nd \log$-form differential equations of arbitrary tree-level cosmological correlators, as we will review them and give the notations of this paper in Sec \ref{sec:not}.
Once we have differential equations, the next step obviously is to solve them. In Sec \ref{sec:1-V} We further introduce the generalized power series expansion method \cite{Moriello:2019yhu} of flat amplitudes to solve this problem. Slightly unlike in \cite{Moriello:2019yhu}, we directly perform just power series expansions (while "generalized" is found to be not necessary here) on the first-order differential equations, rather than deriving the higher-order differential equation for each master integral first. We also find several boundaries whose boundary conditions could be easily determined.
 Surprisingly, due to that the $\nd \log$-form further simplifies the series expansion of differential equations, we find that power series solutions of the vertex integral family exhibit a simple structure, which allows us to conjecture all order expressions of them directly. These solutions are multivariate hypergeometric functions.  We will firstly show two examples in Sec \ref{sec:1-v_1f} and \ref{sec:1-v_2f}: solving the 1-fold and 2-fold Hankel vertex integral families by expanding them around both momentum $k_0$ of the massless leg equals $\infty$ and $k_n$ of one massive leg equals $\infty$, including presenting how to determine their boundary conditions.  The provided power series solutions have a region of convergence. Therefore, in Sec \ref{sec:1V1fcontinue}, we discuss how to perform analytic continuation. The numerical efficiency is also presented in this subsection. In one example, we get the numerical result of points 100 points with a relative error of at most \(\mathcal{O}(10^{-34})\), and evaluating each point only takes about 0.01s by one core of CPU on a personal computer. Then, in Sec \ref{sec:1-v_nf}, we give the constructed all-order power series solution for arbitrary vertex integral family for both boundary $k_0\to\infty$ and $k_n\to\infty$, including their boundary coefficient. These are the new results of this paper.
 Furthermore, in Sec \ref{sec:2-v}, we compute a 2-vertex example. By this example, we want to show that because of the factorization property of IBP  of tree-level cosmological correlators \cite{Chen:2023iix}, these solutions in Sec \ref{sec:1-v_nf} directly give all homogeneous solutions of arbitrary tree-level cosmological correlators.  Although we have not given all solutions of the non-homogeneous part of arbitrary tree-level cosmological correlators, the calculation of the non-homogeneous solution in Sec \ref{sec:2-v} shows using the methods to solve non-homogeneous solutions, including its boundary condition, is also easy and straightforward. Meanwhile, in this section,  we also indicate that blow-up is a useful technique for solving power series expansion of differential equations around degenerate multivariate singularity. The techniques we develop in this paper could also benefit the evaluation of amplitude in flat space, as we also present related discussions in the section of summary and outlook Sec \ref{sec:summary}.
 
 We emphasize that although we have not presented any loop-level example, IBP, differential equations, and generalized power series expansion could be applied to loop-level cosmological correlators straightforwardly as well. Based on our best knowledge, the main possible challenge that could arise at the loop level is determining the boundary conditions. It could be more complicated than tree-level. 
 However, there also are many lessons that can be learned from flat amplitudes for solving boundary conditions. Therefore, it is unlikely to pose a fundamental difficulty.

\section{Background}\label{sec:not}

In this section, we review the basic background for later discussions. 

\subsection {In-in Feynman rules and asymptotic behaviors}

The general Feynman rules for cosmological correlators in in-in formalism are displayed as follows (for a modern review, see \cite{Chen:2017ryl}). The bulk-to-bulk propagators are given by
\begin{align}
G_>(k;\tau_1,\tau_2)\equiv&~u(\tau_1,k)u^*(\tau_2,k),\nn\\
G_<(k;\tau_1,\tau_2)\equiv&~u^{*}(\tau_1,k)u(\tau_2,k).\\
G_{++}(k;\tau_1,\tau_2)=&~G_>(k;\tau_1,\tau_2)\theta(\tau_1-\tau_2)+G_<(k;\tau_1,\tau_2)\theta(\tau_2-\tau_1),\nn\\
G_{+-}(k;\tau_1,\tau_2)=&~G_<(k;\tau_1,\tau_2),\nn\\
G_{-+}(k;\tau_1,\tau_2)=&~G_>(k;\tau_1,\tau_2),\nn\\
G_{--}(k;\tau_1,\tau_2)=&~G_<(k;\tau_1,\tau_2)\theta(\tau_1-\tau_2)+G_>(k;\tau_1,\tau_2)\theta(\tau_2-\tau_1),  \label{eq:proptype}
\end{align}
The mode of the field in the time direction is denoted by $u$, which is
\begin{align}
  u(\tau;k)=-i\frac{\sqrt\pi}{2}e^{i\pi(\nu/2+1/4)}H^{(d-1)/2}(-\tau)^{d/2}\text{H}_{\nu}^{(1)}(-k\tau) \,.
\end{align}
Here the H is Hubble constant, $\text{H}_{\nu}^{(1)}(-k\tau)$ is the Hankel function and other parameters are 
$k=|\bm{k}|$ where $\bm{k}$ is the 3-momentum, $\nu=\sqrt{\frac{d^2}{4}-\frac{m^2}{H^2}}$ and $d=3$.  From this definition, one can see $k$ typically is real and $\nu$ typically is real or imaginary, thus we will usually discuss such cases.
The bulk-to-boundary propagators are non-vanishing only when the field is massless
\begin{align}
G_{+}(k;\tau)\equiv G_{+\pm}(k;\tau_1,0)=\frac{H^2}{2k^3}(1-ik\tau)e^{ik\tau} ,\nn\\
G_{-}(k;\tau)\equiv G_{-\pm}(k;\tau_1,0)=\frac{H^2}{2k^3}(1+ik\tau)e^{-ik\tau} .
\end{align}

The Hankel functions satisfy\footnote{The definition of $\text{H}_{\nu}^{(1,2)}$ in textbook is 
$\text{H}_{\nu}^{(1)}(z)= \text{J}_{\nu}(z)+ i \text{Y}_{\nu}(z)$ and $\text{H}_{\nu}^{(2)}(z)= \text{J}_{\nu}(z)- i \text{Y}_{\nu}(z)$. When $z,\nu$ are real numbers, it is obviously that $\text{H}_{\nu}^{(2)}(z)=(\text{H}_{\nu}^{(1)}(z))^*$. The second line of \eqref{eq:hankel} is a generalization of the above result when $z,\nu$ are complex numbers.} 
\begin{align}
  & \partial_\tau^2 \text{H}_{\nu}^{(1,2)}(-k\tau)+\frac{1}{\tau} \partial_\tau\text{H}_{\nu}^{(1,2)}(-k\tau)+\Big(k^2-\frac{\nu^2}{H^2 \tau^2}\Big)\text{H}_{\nu}^{(1,2)}(-k\tau)=0 \, ,\nn\\
  & \text{H}_{\nu}^{(1)}(-k\tau)= \left(\text{H}_{\nu^\star}^{(2)}(-k^\star\tau^\star) \right)^\star \, .\label{eq:hankel}
\end{align}
The asymptotic behavior of Hankel functions is also important for applying Wick rotation and solving boundary conditions of cosmological correlators. Hence, we list them below. For $\tau\to 0_-$,
\begin{align}
&\text{H}^{(1)}_{\nu}(-k\tau) ~ = c_1(-k\tau)^{v}(1+\mathcal{O}(\tau^2)) ~~ +  c_2(-k\tau)^{-v} (1+\mathcal{O}(\tau^2)) \, , \nn\\
&\text{H}^{(2)}_{\nu^\star}(-k\tau)= c_1^{\star}(-k\tau)^{v^\star}(1+\mathcal{O}(\tau^2)) +  c_2^{\star}(-k\tau)^{-v^\star} (1+\mathcal{O}(\tau^2)) \, , \nn\\
&c_1=e^{-i\pi\nu}c[\nu]\,, ~~ c_2=c[-\nu]\,, ~~
c[\nu]\equiv \frac{ 2^{-\nu } \Gamma (-\nu )}{i \pi }  \, .  \label{eq:Hasy0}
\end{align}
For $k \tau\to -\infty$,
\begin{align}
& \text{H}_{\nu}^{(1)}(-k\tau) \sim \sqrt{\frac{2}{\pi }} (-k \tau)^{-\frac{1}{2}}  e^{-i k \tau-i \pi  \left(\nu/2+1/4\right)}  \, , \nn\\
& \text{H}_{\nu}^{(2)}(-k\tau) \sim \sqrt{\frac{2}{\pi }} (-k \tau)^{-\frac{1}{2}}  e^{i k \tau+i \pi  \left(\nu/2+1/4\right)}  \, . \label{eq:Hasyinf}
\end{align}

\subsection{Notations for indices}

Since we will frequently use the tensor product of 2-component vectors, we also introduce the following notation for convenience: 
\begin{align}
&\bm{a}\equiv a_1,a_2,\cdots,a_n  \,~, ~~~~  a_i = 0,1 \, ,\nn\\
&\tilde{\bm{a}}\equiv1+\sum_{i=1}^n a_i 2^{n-i}  \,, \nn\\
&\text{I}_{\tilde{\bm{a}}}\equiv \text{I}_{\{\bm{a}\}}=\text{I}_{\{a_1,a_2,\cdots\}}\label{eq:defindex}
\end{align}
Let us explain the meaning of the above notations. The $\bm{a}$ is a vector with $n$ components, while each component takes only two values $0$ and $1$. Thus we can write $\bm{a}$ as a binary, for example, $\bm{a}=10100$. The meaning of $\tilde{\bm{a}}$ is to transfer 
the binary number $\bm{a}$ to a number in decimal system, for example, $\tilde{0}=1$, $\tilde{1}=2$ and $\widetilde{1010}=11$. 
In other words, we should treat the $~\widetilde{}~$ as an operation acting on $\bm{a}$. Using this action, we can easily get
the location of $\bm{a}$-th component in the tensor product. 
For instance, in the 2-fold vertex integral family, there are four master integrals. The $3$-th master integrals can be denoted as $\text{I}_{3}$ or $\text{I}_{\{1,0\}}$. Another example is that   $f_{\{a_3=1\}}=f_{\tilde{a}_3=2}$. 


\subsection{Integral family and differential equations}

For in-in Feynman diagrams, we can use the following elements to express integrands of tree diagrams
\begin{align}
& {\cal{I}} = \int_{-\infty}^0  \prod_i \nd \tau_i \tau_i^{\alpha_i} \prod_j F_j\, , 
\nn\\
 & F_j  = e^{i k \tau} 
, ~~ \text{H}_{\nu}^{(1,2)} (-k \tau), ~~ \partial_{\tau} \text{H}_{\nu}^{(1,2)}(-k \tau), ~~ \theta(\tau_j-\tau_k) \, . \label{eq:dSintegrand}
\end{align}
Here, each $\int \nd \tau_i$ corresponds to a time-integration of a vertex. In this paper, we will denote the tree-level cosmological correlators with $M$ vertices as ``$M$-vertex correlators", including integrals with respect to $M$ time variables $\tau_i$. We call the integral family with one vertex a ``vertex integral family". Since each massive leg contributes a Hankel function in the integrand, we use the $n$-fold (Hankel) vertex integral family to denote a vertex integral family with $n$ massive legs. Integrals in the family can be written as
\begin{align}
&f^{(a_0)}_{\tilde{\bm{a}},\bm{s}}=\int_{-\infty}^0 (-\tau)^{\nu_0+a_0} e^{ik_0 \tau} 
  \prod_{i=1}^n h^{(s_i)}(\nu_i,a_i;-k_i\tau)\nd \tau \,, \nn\\
&~~~a_0\in \mathbb{Z}\, , ~~ s_{i\geq 1}, a_{i\geq 1}=0,1;~~\bm{a}=(a_1,\cdots,a_n);~~\bm{s}=(s_1,...,s_n)
\end{align}
where $h$-functions are redefined using the Hankel function and its time derivative for  later convenience: 
\begin{align}
 &h^{(1\text{ or }2)}(\nu,0;-k\tau)\equiv  (-k\tau)^{-\nu} \text{H}_{\nu}^{(1)}(-k \tau) ~ (\text{or }\text{H}_{\nu^\star}^{(2)} )\propto \tau^{-\frac{3}{2}-\nu} u ~ (\text{or } u^*),\nn\\
& h^{(1\text{ or }2)}(\nu,1;-k\tau)\equiv - \frac{1}{k} \partial_\tau h^{(1\text{ or }2)}(\nu,0;-k\tau),
\end{align}
The iterative IBP reduction and $\nd \log$-form differential equations have been given in \cite{Chen:2023iix}. It says $n$-fold vertex integral family has $2^n$ master integrals, which we 
denote as $\text{I}_{\tilde{\bm{a}}}$. 
If one selects all $\text{I}_{\tilde{\bm{a}}}= f^{(0)}_{\tilde{\bm{a}},\bm{s}}$\footnote{Although the function $f^{(0)}$ depends on the choice of $\bm{s}$ (so is $\text{I}_{\tilde{\bm{a}}}$), the matrix $(\nd \Omega)$ given in \eref{eq:ndlogDE} does not depend on $\bm{s}$ \cite{Chen:2023iix}. 
 }, with $a_i=0,1$, as  master integrals, the differential equations of them are automatically $\nd \log$-form and given by the uniform formula:
\begin{align}
&\nd \text{I} = (\nd \Omega) \cdot \text{I}  = \sum_{i=0}^{n} \Omega_{k_i} \cdot \text{I}  ~~\nd k_i  \, ,\nn\\
&\Omega =  \Omega_{ex} - i \text{T}_n^{-1}\cdot  \tilde{\Omega}_0 \cdot \text{T}_n \cdot \text{M}_1[\nu_0+1,\bm{\nu}] \, .\label{eq:ndlogDE}
\end{align}
where $\Omega_{k_i}={\partial \over \partial k_i}\Omega $
with
\ie \label{eq:omega-form}
&\left( \tilde{\Omega}_0 \right)_{\bm{b}\bm{a}}  \equiv \begin{cases} - i \log \Big[ k_0+\sum_i (2a_i-1)k_i\Big] , & \bm{b}=\bm{a} \\
0, & \bm{b}\neq \bm{a} 
\end{cases} \, , \\
&\left( \Omega_{ex} \right)_{\bm{b}\bm{a}}  \equiv \begin{cases} - \sum_i a_i(2\nu_i+1) \log k_i , & \bm{b}=\bm{a} \\
0, & \bm{b}\neq \bm{a} 
\end{cases} \, ,
\fe
and
\ie \label{eq:M-form}
&~~\left(\text{M}_1[\nu_0,\bm{\nu}]\right)_{\bm{b}\bm{a}}= \begin{cases}
\nu_0-\sum_i a_i(2\nu_i+1) , & \bm{b}=\bm{a} \\
0, & \bm{b}\neq \bm{a}\end{cases}\, ,\\
&~~\left(\text{T}_n\right)_{\bm{b}\bm{a}}=\prod_{i=1}^n \text{T}_{b_i a_i}  \, ,
~~~~~ \text{T}=\frac{1}{\sqrt{2}}\left(
\begin{array}{cc}
 1 & -i \\
 -i & 1 \\
\end{array}
\right)  \, .
\fe

For $M$-vertex correlators, using 2-vertex correlators as an example for simple, besides the master integrals coming from the product of two 1-vertex master integrals, there will be an extra master integral coming from the remaining term of IBP which appears due to the step functions when applying the IBP method. More discussions can be found in \cite{Chen:2023iix}. For propagator $G_{\pm\mp}$, the master integrals are just the product of two 1-vertex master integrals. For example, if the propagator is $G_{+-}(\tau_1,\tau_2)$, we have 
\ie
{\cal{I}}_{\tilde{\bm{r}};+-} \equiv& 
\left(\int \nd \tau_1 \hat{\text{I}}^{(0)}_{\bm{a},\bm{s}_1;1}\right)
\left(\int\nd \tau_2 \hat{\text{I}}^{(0)}_{\bm{b},\bm{s}_1;2}\right),~~~\bm{r}=\bm{a},\bm{b} 
\fe
with
\ie
\hat{\text{I}}^{(0)}_{\bm{a},\bm{s}_1;1} = (-\tau_1)^{\nu_{0;1}} e^{ik_{0;1}\tau_1}\prod_i   h^{(s_i)}(\nu_{i;1},a_i,-k_{i;1}\tau_1).
\fe
For propagator $G_{\pm\pm}$, things are more complicated due to the step function, and the remaining term will appear. In the $G_{++}$ case, the selected master integrals and the $\nd\log$-form are\footnote{When we integrate the delta-function, two $h$ will combine to give a simple factor (see Eq(3.11) and Eq(3.12) of \cite{Chen:2023iix} ), so for remaning part, we have $r=\bm{a}_{\hat{i}},\bm{b}_{\hat{j}}$ as given in the second line in \eref{eq:2vmide}. The matrix $\Omega_{\tilde{\bm{r}}\tilde{\bm{s}};++}$ gives the reduction coefficients of the differential of the original sector to
the remaining part. More explanation can be found in Eq.(3.68)  of \cite{Chen:2023iix}. }
\ie \label{eq:2vmide}
{\cal{I}}_{\tilde{\bm{r}};++} \equiv& \int \nd \tau_1 \nd \tau_2  \hat{\text{I}}^{(0)}_{\bm{a},\bm{s}_1;1} \theta_{1,2}^{(i,j)} \hat{\text{I}}^{(0)}_{\bm{b},\bm{s}_2;2} \, ,~~~~ \bm{r}=\bm{a},\bm{b}   \, ,\\
\bm{R}_{\tilde{\bm{r}};++}=&-\delta_{a_i,1-b_j} (-1)^{a_i+1} \frac{4i}{\pi} e^{ \pi \text{Im}[\nu]} (k_{i;1})^{-2\nu_{i;1}-1} f^{(-2\nu_{i;1})}_{\bm{a}_{\hat{i}},\bm{b}_{\hat{j}} } \, ,~~~~ \bm{r}=\bm{a}_{\hat{i}},\bm{b}_{\hat{j}}  \, ,\\
\Omega_{\tilde{\bm{r}}\tilde{\bm{s}};++} = &- i \left( \text{T}_n^{-1}.  \tilde{\Omega}_{0;1}.\text{T}_n \right)_{\bm{a}(\bm{c}_{\hat i};1-b_j)} \delta_{b_{\hat j}d_{\hat j}} (-1)^{b_j}  \\
& -  i \left( \text{T}_n^{-1}.  \tilde{\Omega}_{0;2}.\text{T}_n \right)_{\bm{b}(\bm{d}_{\hat i};1-a_i)} \delta_{a_{\hat i}c_{\hat i}} (-1)^{a_i}
\, ,~~~~ \bm{r}=\bm{a},\bm{b};~~\bm{s}=\bm{c}_{\hat{i}},\bm{d}_{\hat{j}}  \, .
\fe
We have used the following notation in the expression above:
\ie
&h(\nu_{i;1},a_i,-k_{i;1}\tau_1) \theta_{1,2}^{(i,j)} h(\nu_{j;2},b_j,-k_{j;2}\tau_2)\equiv  \nn\\
&h^{(1)}(\nu_{i;1},a_i,-k_{i;1}\tau_1) \theta_{12} h^{(2)}(\nu_{j;2},b_j,-k_{j;2}\tau_2) + h^{(2)}(\nu_{i;1},a_i,-k_{i;1}\tau_1) \theta_{21} h^{(1)}(\nu_{j;2},b_j,-k_{j;2}\tau_2) \, , \nn\\
&\nu_{i;1}= \nu_{j;2} \,, \ \ k_{i;1}=k_{j;2}\,,~~~~\theta_{ij}=\theta(\tau_i-\tau_j) \, 
\fe
 This paper will only discuss the $G_{++}$ case of 2-vertex correlators and hence we will suppress the label $++$ in \eqref{eq:2vmide} when we consider the 2-vertex case in Sec \ref{sec:2-v}.

To express the solutions to these differential equations more compactly, we also use the {\bf Pochhammer symbol $(a)_n\equiv \Gamma(a+n)/\Gamma(a)$} in some expressions. 

\section{Analytic results of $n$-fold vertex integral family}\label{sec:1-V}

In this section, we will derive the analytic expression for a single vertex with $n$ Hankel functions. 

\subsection{Pedagogical example: 1-fold vertex integral family}\label{sec:1-v_1f}

\subsubsection{Preparation}
As a pedagogical example, let us solve the 1-fold Hankel vertex integral family\footnote{For simplicity, we write
$ h(\nu,a,-k_1\tau)= h_{\nu}(a,-k_1\tau)$. }:
\begin{align}
\int_{-\infty}^0 \nd \tau e^{ i k_0 \tau} (-\tau)^{\nu_0+a_0} h^{(2)}_{\nu_1}(a,-k_1\tau) \, .
\end{align}
To simplify the discussion, we will consider $\nu_i \in \mathbb{R}$ in this section. The methods employed here can be applied directly to the case of imaginary values of $\nu_i$ as well. We will also discuss how to extend the results to imaginary $\nu_i$ in Sec \ref{sec:1V1fcontinue}.

This function family has 2 master integrals. We use following $\text{I}_{\tilde{a}}$ as master integrals:
\begin{align}
&\text{I}_{1}=\int_{-\infty}^0 \nd \tau e^{ i k_0 \tau} (-\tau)^{\nu_0} h^{(2)}_{\nu_1}(0,-k_1\tau) \,,\nn\\ 
&\text{I}_{2}=\int_{-\infty}^0 \nd \tau e^{ i k_0 \tau} (-\tau)^{\nu_0} h^{(2)}_{\nu_1}(1,-k_1\tau)\,.
\end{align}
This choice of master integrals is the same as those constructed in \cite{Chen:2023iix}, allowing us to directly use the uniform formula of differential equations of them \eqref{eq:ndlogDE}. The differential equations are as follows:
\begin{align}
&\nd \text{I}_{\tilde{a}} = \left (\nd \Omega_{\tilde{a}\tilde{b}} \right) \text{I}_{\tilde{b} }  \,,\nn\\
&\Omega_{11}=
 -i \left(\nu _0+1\right) \left(-\frac{1}{2} i \log \left(k_0-k_1\right)-\frac{1}{2} i \log
   \left(k_0+k_1\right)\right)  \,,\nn\\
&\Omega_{12}=-i \left(\nu _0-2 \nu _1\right) \left(\frac{1}{2} \log
   \left(k_0+k_1\right)-\frac{1}{2} \log \left(k_0-k_1\right)\right)   \,, \nn\\
&\Omega_{21}=
 -i \left(\nu _0+1\right) \left(\frac{1}{2} \log \left(k_0-k_1\right)-\frac{1}{2} \log
   \left(k_0+k_1\right)\right)   \,, \nn\\
&\Omega_{22}=-\left(2 \nu _1+1\right) \log \left(k_1\right)-i \left(\nu _0-2 \nu _1\right)
   \left(-\frac{1}{2} i \log \left(k_0-k_1\right)-\frac{1}{2} i \log \left(k_0+k_1\right)\right)   \,.  \label{eq:dlogDE1V1f}
\end{align}
This is a system of first-order differential equations. It is well known that, for first-order differential equations, one can determine the solution if the values of the principal integral on the boundary or its asymptotic behavior are provided. Similar to the application of differential equation methods in flat spacetime field theory, a natural approach here is to choose a boundary point that is significantly easier to compute than the original integral. Noting that the Wick rotation of this integral family is 
\begin{equation}
\tau \to - i \tau
\end{equation} 
and the asymptotic behavior \eqref{eq:Hasyinf} of Hankel functions at $k\tau\to -\infty$, taking the limit as $k_i\to \infty$ (since $\tau \leq 0$) simplifies the computation due to the exponential suppression. We will proceed with calculations using this boundary.

In the following part of this section, we will first solve the system of differential equations using the method of power series expansion around $k_0\to \infty$. We will then find this analytical series solution can be rewritten as hypergeometric functions, thereby obtaining a compact analytical function. Subsequently, we will determine the coefficients of the analytical solution by computing the boundary conditions.

\subsubsection{Solutions with the boundary $k_0\to \infty$} \label{sec:1v1fk0inf}
For convenience, we define a new parameter 
\begin{equation}
x=\frac{1}{k_0}.
\end{equation} 
Then, the matrix of partial differential equations with respect to $x$ is
\begin{align}
&\partial_x \text{I}_{\tilde{a}} = \left(\Omega_x\right)_{\tilde{a} \tilde{b}} \text{I}_{\tilde{b}}   \, , ~~~~~~\Omega_x = \left( 
\begin{Large}
\begin{array}{cc}
 \frac{\nu _0+1}{x-k_1^2 x^3} & \frac{i k_1 \left(\nu _0-2 \nu _1\right)}{k_1^2 x^2-1} \\
 -\frac{i k_1 \left(\nu _0+1\right)}{k_1^2 x^2-1} & \frac{\nu _0-2 \nu _1}{x-k_1^2 x^3} \\
\end{array}
\end{Large}
\right)  \,.  \label{eq:DEx1V1fk0inf}
\end{align} 
Due to the differential equation of the chosen master integrals being in  $\nd \log$-form, its series expansion takes a very simple form. The power series expansion of $\Omega_x$ around $x = 0$ are
\begin{align}
&\Omega_x = \sum_{i=-1}^\infty \Omega_x^{(i)} x^{i} \,, \nn\\
&\Omega_x^{(-1+2j)}=\left(
\begin{array}{cc}
 \left(\nu _0+1\right) k_1^{2j} & 0 \\
 0 &\left(\nu_0-2 \nu _1\right) k_1^{2j} \\
\end{array}
\right)     \,,\nn\\
&\Omega_x^{(0+2j)}=\left(
\begin{array}{cc}
 0 & -i  \left(\nu _0-2 \nu _1\right)k_1^{1+2j} \\
 i \left(\nu _0+1\right) k_1^{1+2j} & 0 \\
\end{array}
\right) \,, \label{eq:1vexpansionDE}
\end{align}
where  $j\in \mathbb{N}$.
We also denote the ansatz of the power series expansions of the solutions around $x=0$ as
\begin{align}
f_{i}=x^\lambda \sum_{j=0}^\infty \text{C}(i,j) x^j\,
\end{align} 
where $\lambda$ represents the smallest nonzero exponent among all master integrals of this solution. To determine $\lambda$, we consider the indicial equation derived from the leading order of the power series solution of the differential equations:
\begin{align}
& ~~~~~~~~~ \partial_x \text{C}(i,0) x^\lambda = \left(\Omega_x^{(-1)}\right)_{ij} x^{-1} \text{C}(j,0) x^\lambda \nn\\
& \Rightarrow \lambda \left(
\begin{array}{c}
\text{C}(1,0) \\
 \text{C}(2,0) \\
\end{array}
\right) =  \left(
\begin{array}{cc}
 \nu_0+1 & 0 \\
 0 & \nu_0-2 \nu _1 \\
\end{array}
\right) . \left(
\begin{array}{c}
\text{C}(1,0) \\
 \text{C}(2,0) \\
\end{array}
\right)
\end{align} 
Solving for $\text{C}(1,0)$ and $\lambda$ yields two non-trivial solutions. They are
\begin{align}
&\text{solution 1: }~\lambda=\nu_0+1 \,,~~~~~~\text{C}(2,0)=0 \,, \nn\\
&\text{solution 2: }~\lambda=\nu_0-2\nu_1 \,,~~~\text{C}(1,0)=0 \,.
\end{align} 
For each selected solution, one can solve $\text{C}(i,j)$ iteratively. For example, the $x^{\lambda+j_0}$ order of \eqref{eq:DEx1V1fk0inf} gives
\begin{align}
(\lambda+j_0)\text{C}(i,j_0)=\sum_{j=-1}^{j_0} \Omega_x^{(j)}.~\text{C}(i,j_0-j) \,. \label{eq:seriesEqs}
\end{align}
Supposing people have solved $\text{C}(i,j<j_0)$, people can solve $\text{C}(i,j_0)$ from the above equations. As a result, the master integrals could be expressed as
\begin{align}
\text{I}_{\tilde{\bm{a}}}= \text{C}^{[1]}f_{\tilde{\bm{a}}}^{[1]}+ \text{C}^{[2]}f_{\tilde{\bm{a}}}^{[2]} \,,  \label{eq:1V1fsumsol}
\end{align}
where referring to the definition of index $\tilde{\bm{a}}$ in \eqref{eq:defindex}, $\text{I}_{\tilde{\bm{a}}}$ denotes the $\tilde{\bm{a}}$-th master integral.
 Here we denote the $i$-th function in the $j$-th general solution of the differential equations by $f_i^{[j]}$. We denote the boundary coefficients $\text{C}(1,0)$ in solution 1 by $\text{C}^{[1]}$ and the $\text{C}(2,0)$ in solution 2 by $\text{C}^{[2]}$, which will be determined by boundary conditions. $f_i^{[j]}$ and  $\text{C}^{[j]}$ together give the particular solutions corresponding to master integrals. 
The power series solutions $f_i^{[j]}$ of differential equations in \eqref{eq:1V1fsumsol} are
\begin{align}
&f_1^{[1]}(1/x,k_1) = x^{\nu_0+1} \sum_{m=0}^\infty \frac{ \left(\frac{\nu_0+1}{2}\right)_m \left(\frac{\nu_0+2}{2}\right)_m   }{(\nu_1+1)_m }\frac{(k_1^2 x^2)^m}{m!} \,, \nn\\
&f_2^{[1]}(1/x,k_1)= x^{\nu_0+1}
i k_1 x \sum_{m=0}^\infty \frac{ \left(\frac{\nu_0+1}{2}\right)_{m+1} \left(\frac{\nu_0+2}{2}\right)_m   }{(\nu_1+1)_{m+1} }\frac{(k_1^2 x^2)^m}{m!}  \,, \nn\\
&f_1^{[2]}(1/x,k_1)= x^{\nu_0-2\nu_1} (-i k_1 x)\sum_{m=0}^\infty \frac{ \left(\frac{\nu_0-2\nu_1}{2}\right)_{m+1} \left(\frac{\nu_0-2\nu_1+1}{2}\right)_m   }{(-\nu_1)_{m+1} }\frac{(k_1^2 x^2)^m}{m!} \,, \nn\\
&f_2^{[2]}(1/x,k_1)  = x^{\nu_0-2\nu_1}\sum_{m=0}^\infty \frac{ \left(\frac{\nu_0-2\nu_1}{2}\right)_m \left(\frac{\nu_0-2\nu_1+1}{2}\right)_m   }{(-\nu_1)_m }\frac{(k_1^2 x^2)^m}{m!} \,, \label{eq:1V1fsolseriesk0inf}
\end{align}
which converge when $|k_1/k_0|<1$.
Differential equations being in $\nd \log$-form leads to the simple expression of the expansion of differential equations and the power series solution as well. As a result, we can easily see that these solutions can be expressed in terms of well-known hypergeometric functions: 
\begin{align}
&f_1^{[1]}(1/x,k_1) = x^{\nu_0+1} \, _2\text{F}_1\left(\frac{\nu _0+1}{2},\frac{\nu_0+2}{2};\nu _1+1;k_1^2 x^2\right) \,, \nn\\
&f_2^{[1]}(1/x,k_1)= x^{\nu_0+1}
\frac{i k_1 \left(\nu _0+1\right) x }{2 \left(\nu _1+1\right)} \,  _2\text{F}_1\left(\frac{\nu_0+2}{2},\frac{\nu_0+3}{2};\nu_1+2; k_1^2 x^2\right)  \,, \nn\\
&f_1^{[2]}(1/x,k_1)= x^{\nu_0-2\nu_1} \frac{i k_1 \left(\nu _0-2 \nu _1\right) x }{2 \nu _1} 
 \, _2\text{F}_1\left(\frac{\nu _0-2\nu_1+1}{2},\frac{\nu _0-2\nu_1+2}{2};1-\nu _1;  k_1^2 x^2\right) \,, \nn\\
&f_2^{[2]}(1/x,k_1)  = x^{\nu_0-2\nu_1} \, _2\text{F}_1\left(\frac{\nu _0-2\nu_1}{2},\frac{\nu _0-2\nu_1+1}{2};-\nu _1;k_1^2 x^2\right) \,. \label{eq:1V1fsolF-1}
\end{align}
 To complete the calculation of the master integrals, we only need to determine the coefficients $\text{C}^{[1]}$ and $\text{C}^{[2]}$ by boundary conditions as follows.

Due to the exponential factor \( e^{ik\tau} \) being suppressed when \( k_0 \to \infty \) after a Wick rotation, only the region where \( \tau \to 0 \) can contribute non-zero terms. Therefore, we expand the other parts of the integrand around \( \tau = 0 \).
Recall \eqref{eq:Hasy0}, we have
\begin{align}
&h_\nu^{(1)}(0,-k\tau)=c_1 (1+\mathcal{O}(\tau^2)) +  c_2 (-k\tau)^{-2\nu} (1+\mathcal{O}(\tau^2)) \, , \nn\\
&h_\nu^{(2)}(0,-k\tau)=c_1^{\star} (-k\tau)^{-\nu+\nu^\star}(1+\mathcal{O}(\tau^2)) +  c_2^{\star} (-k\tau)^{-\nu-\nu^\star} (1+\mathcal{O}(\tau^2)) \, . \label{eq:asyhRInu}
\end{align}
We denote the coefficient we need for boundary condition by $\text{C}_{\tilde{\bm{a}}}^{(k_0)}$.
For real $\nu$ and $h^{(2)}_\nu(a,-k\tau)$, the case we consider in this section, we have
\begin{align}
&h_{\nu}^{(2)}(0,-k\tau)\sim c_1^\star +  c_2^\star (-k\tau)^{-2\nu}  = \text{C}_1^{(k_0)}(\nu)  +  \mathcal{O}(\tau^{-2\nu})   \, , \nn\\
&h_{\nu}^{(2)}(1,-k\tau)\sim-2\nu c_2^\star (-k\tau)^{-2\nu-1}=\text{C}_2^{(k_0)}(\nu) (-k\tau)^{-2\nu-1}\, ,\nn\\
&\text{C}_1^{(k_0)}(\nu) = c_1^\star=-e^{i\pi\nu}c[\nu] \, , ~~\text{C}_2^{(k_0)}(\nu) = -2\nu c_2^\star=c[-\nu-1] \, , \label{eq:expandha2R}
\end{align}
where $c[\nu]$ is defined in  \eqref{eq:Hasy0}. The $\mathcal{O}(\tau^{-2\nu})$ term in $h_\nu^{(2)}(0,-k\tau)$ contributes to next-to-leading-order term of solution two. Hence we do not need its coefficient here.
Taking the expansions in the integrands,  we have 
\begin{align}
& \text{C}^{[1]}=k_0^{\nu_0+1}\int_{-\infty}^0 \nd \tau e^{ i k_0 \tau} (-\tau)^{\nu_0} c_{1} = (-i)^{\nu_0+1} \Gamma \left(\nu_0+1\right) \text{C}_1^{(k_0)}(\nu_1) \, , \nn\\
& \text{C}^{[2]}=k_0^{\nu_0-2\nu_1}\int_{-\infty}^0 \nd \tau e^{ i k_0 \tau} (-\tau)^{\nu_0} (-2 \nu_1 c_{2})(-k_1 \tau)^{-2\nu_1-1} =(-i)^{\nu_0-2\nu_1}  \Gamma \left(\nu _0-2 \nu _1\right) \frac{ \text{C}_2^{(k_0)}(\nu_1)}{k_1^{2 \nu _1+1}} \,.
\end{align}
After substituting this result into \eqref{eq:1V1fsumsol},  the calculation is completed.

\subsubsection{Solutions with the boundary $k_1\to \infty$}\label{sec:1v1fk1inf}

In this section, we will present an alternative boundary condition choice $k_1\to \infty$. The series solution obtained under this choice can cover the region uncovered by the convergence region of the series solution from the previous section.
Although \eqref{eq:1V1fsolF-1} allows us to directly obtain the power series solutions, this approach is not feasible for more complex cases. Therefore, as an instructional case, again we derive the series solution through a differential equation expansion.
We define
\begin{equation}
x=\frac{1}{k_1}.
\end{equation} 
Then, the differential equations are
\begin{align}
&\Omega_x =\left(
\begin{array}{cc}
 \frac{\nu _0+1}{x-k_0^2 x^3} & \frac{i k_0 \left(\nu _0-2 \nu _1\right)}{k_0^2 x^2-1} \\
 -\frac{i k_0 \left(\nu _0+1\right)}{k_0^2 x^2-1} & \frac{k_0^2 \left(2 \nu _1+1\right) x^2-\nu _0-1}{x
   \left(k_0^2 x^2-1\right)} \\
\end{array}
\right)  \,.  \label{eq:DEx1V1fk1inf}
\end{align} 
The series expansion of $\Omega_x$ is
\begin{align}
&\Omega_x = \sum_{i=-1}^\infty \Omega_x^{(i)} x^{i} \,, \nn\\
&\Omega_x^{(-1)}=\left(
\begin{array}{cc}
 \left(\nu _0+1\right) & 0 \\
 0 &\left(\nu_0+1\right) \\
\end{array}
\right)     \,,\nn\\
&\Omega_x^{(0+2j)}=\left(
\begin{array}{cc}
 0 & i  \left(\nu _0+1\right)k_0^{1+2j} \\
 -i \left(\nu_0-2\nu_1\right) k_0^{1+2j} & 0 \\
\end{array}
\right) \,, \nn\\
&\Omega_x^{(1+2j)}=\left(
\begin{array}{cc}
 \left(\nu _0+1\right) k_0^{2j+2} & 0 \\
 0 &\left(\nu_0-2\nu_1\right) k_0^{2j+2} \\
\end{array}
\right)     \,,
\end{align}
$\Omega_x^{(-1)}$ gives indicial equations and the solution is
\begin{align}
\lambda=\nu_0+1 \,.
\end{align}
Since $\text{C}(1,0)$ and $\text{C}(2,0)$ are undetermined, the system of differential equations still has two linear independent solutions and corresponding two boundary coefficients $\text{C}^{[1]}$ and $\text{C}^{[2]}$.
Then, we solve the equations at each order of $x$ again and find the power series solutions are
\begin{align}
&f_1^{[1]}(k_0,1/x) =x^{\nu_0+1} \sum_{m=0}^\infty  \left(\frac{\nu _0+1}{2}\right)_m \left(\frac{\nu _0-2 \nu _1+1}{2}\right)_m \frac{4^m \left(k_0 x\right)^{2 m} }{(2 m)!} \,, \nn\\
&f_2^{[1]}(k_0,1/x) = x^{\nu_0+1}
i k_0 x (\nu_0+1)  \sum_{m=0}^\infty   \left(\frac{\nu _0+3}{2}\right)_m \left(\frac{\nu _0-2 \nu _1+1}{2}\right)_m \frac{4^m \left(k_0 x\right)^{2 m} }{(2 m+1)!}  \,, \nn\\
&f_1^{[2]}(k_0,1/x)= x^{\nu_0+1} (-i k_0 x) (\nu_0-2\nu_1) \sum_{m=0}^\infty  \left(\frac{\nu _0+2}{2}\right)_m \left(\frac{\nu _0-2 \nu _1+2}{2}\right)_m \frac{4^m \left(k_0 x\right)^{2 m} }{(2 m+1)!} \,, \nn\\
&f_2^{[2]}(k_0,1/x)  = x^{\nu_0+1} \sum_{m=0}^\infty  \left(\frac{\nu _0+2}{2}\right)_m \left(\frac{\nu _0-2 \nu _1}{2}\right)_m \frac{4^m \left(k_0 x\right)^{2 m} }{(2 m)!}    \,, \label{eq:1V1fsolserieskninf}
\end{align}
which converge when $|k_0/k_1|<1$. They also could be expressed as hypergeometric functions:
\begin{align}
&f_1^{[1]}(k_0,1/x) = x^{\nu_0+1} \, _2\text{F}_1\left(\frac{\nu _0+1}{2},\frac{\nu _0-2\nu_1+1}{2};\frac{1}{2};k_0^2 x^2\right) \,, \nn\\
&f_2^{[1]}(k_0,1/x) = x^{\nu_0+1}
i k_0 x (\nu_0+1) \,  _2\text{F}_1\left(\frac{\nu_0+3}{2},\frac{\nu _0-2\nu_1+1}{2};\frac{3}{2};k_0^2 x^2\right)  \,, \nn\\
&f_1^{[2]}(k_0,1/x) = x^{\nu_0+1}(-i k_0 x) (\nu_0+1) 
 \, _2\text{F}_1\left(\frac{\nu_0+2}{2},\frac{\nu _0-2\nu_1+2}{2};\frac{3}{2};k_0^2 x^2\right) \,, \nn\\
&f_2^{[2]}(k_0,1/x)  = x^{\nu_0+1} \, _2\text{F}_1\left(\frac{\nu_0+2}{2},\frac{\nu_0-2\nu_1}{2};\frac{1}{2};k_0^2 x^2\right) \,. \label{eq:1V1fsolF}
\end{align}
Expanding \eqref{eq:1V1fsolF} at $k_1 \to \infty$, we have 
\begin{align}
&\int_{-\infty}^0 \nd \tau e^{ i k_0 \tau} (-\tau)^{\nu_0} h^{(2)}_{\nu_1}(0,-k_1\tau)=\text{C}_1^{(k_1)} x^{\nu_0+1} + \mathcal{O}(x^{\nu_0+2}) \, , \nn\\
&\int_{-\infty}^0 \nd \tau e^{ i k_0 \tau} (-\tau)^{\nu_0} h^{(2)}_{\nu_1}(1,-k_1\tau)=\text{C}_2^{(k_1)} x^{\nu_0+1} + \mathcal{O}(x^{\nu_0+2}) \, , \nn\\
&\text{C}_1^{(k_1)}(\nu_0,\nu_1) =\frac{\pi  2^{\nu _0-\nu _1+2} e^{i \pi \nu_0/2 } }
{ \left(1+e^{i \pi  \nu _0}\right)
\left(1+e^{i \pi  \left(\nu _0-2 \nu _1\right)}\right) 
\Gamma \left(\frac{1}{2}-\frac{\nu _0}{2}\right)
\Gamma\left(-\frac{\nu _0}{2}+\nu _1+\frac{1}{2}\right)}  \,,\nn\\
&\text{C}_2^{(k_1)}(\nu_0,\nu_1) =-\frac{i \pi  2^{\nu _0-\nu _1+2} e^{i \pi \nu_0 /2} }
{\left(-1+e^{i \pi  \nu _0}\right)
\left(-1+e^{i \pi  \left(\nu _0-2 \nu _1\right)}\right) 
\Gamma \left(-\frac{\nu _0}{2}\right) 
\Gamma\left(-\frac{\nu _0}{2}+\nu _1+1\right)} \, , \label{eq:1V1fBCCk1inf}
\end{align}
and
\begin{align}
& \text{C}^{[1]}=\text{C}_1^{(k_1)}(\nu_0,\nu_1) \,,~~~ \text{C}^{[2]}=\text{C}_2^{(k_1)}(\nu_0,\nu_1)      \, . 
\end{align}
In this section, obtaining \eqref{eq:1V1fBCCk1inf} seems like circular reasoning. However, \eqref{eq:1V1fBCCk1inf} will assist us in determining the boundary conditions for more complex cases.

\subsubsection{Analytic Continuation and Numerical Computation Efficiency} \label{sec:1V1fcontinue}

Note that the series solutions obtained in previous sections have a finite region of convergence. This happens commonly when using the series expansion method. 
Although we may be able to solve the series solution in another region like we have done in Sec \ref{sec:1v1fk1inf}, it could not be easy in general cases. Hence, in this section, we will discuss how to extend the series solution to regions outside its convergence domain in the general case. We will continue to use the 1-fold vertex integral family as an example for clarity in some places.

We will outline two methods for analytic continuation.
Firstly, in the 1-fold vertex integral family we solved, we expressed the analytical series result in terms of a known hypergeometric function, whose properties are well-studied. One can directly use Gauss inverse relation
\begin{align}
_2\text{F}_1\left(a,b;c;z\right) &=  \frac{\Gamma(c)\Gamma(b-a)}{\Gamma(b)\Gamma(c-a)} (-z)^{-a} _2\text{F}_1\left(a,1+a-c;1+a-b;\frac{1}{z}\right) \nn\\
&~ + \frac{\Gamma(c)\Gamma(a-b)}{\Gamma(a)\Gamma(c-b)} (-z)^{-b} _2\text{F}_1\left(b,1+b-c;1+b-a;\frac{1}{z}\right) \,
\end{align}
to get the power series around $z=\infty$. For general cases, we may obtain a series of solutions corresponding to generalized hypergeometric functions with multi-variables, some properties of such functions can be found in \cite{Bailey1935GeneralizedHS,Slater1966GeneralizedHF,Exton1979HandbookOH,Exton1979MultipleHF,Srivastava1985MultipleGH}. However, extending these series of solutions beyond the radius of convergence may not always have a ready-made result. One possible method for achieving such analytic continuation is through the Mellin-Barnes contour \cite{barnes1908new,WhittakerACO}. This scenario commonly appears in systems involving IBP and differential equations, and the calculations in flat spacetime QFT have driven related research  \cite{Feng:2024xio}. Secondly, since there is no fully understood analytic function to represent the integral we need to compute, one might consider defining new "analytic functions" directly through differential equations \cite{Liu:2023jkr}. With differential equations, we can easily solve a series expansion solution at any point of parameter space and obtain extremely accurate numerical results with remarkable speed \cite{Moriello:2019yhu}. Automatic packages for numerical differential equations \cite{Hidding:2020ytt,Liu:2022chg} are already available and widely used in flat QFT.  This means that one can quickly compute function values at any regular point and analyze the asymptotic behavior near any singularities. Then, as long as boundary conditions are given, these functions appear to have not many differences from a so-called "analytic" result.

To illustrate, let us assume that we do not know the properties of the hypergeometric function in \eqref{eq:1V1fsolF}, and only have the series solution given by \eqref{eq:1V1fsolseriesk0inf} with its domain of convergence to be \(|k_1 / k_0| < 1\). We will demonstrate the second method to obtain function values where \(|k_1 / k_0| > 1\).

The numerical method is straightforward. We begin by obtaining the function value at a point within the convergence domain \(|k_1 / k_0| < 1\). This value is then used as a new boundary condition to expand.  Solving the linear system like \eqref{eq:seriesEqs} again provides the function value at another point. By selecting a series of points to form a path that bypasses the singularities, we can extend the solution to region \(|k_1 / k_0| > 1\). For example, consider \(v_0 = 54/5\) and \(v_1=11/7\). We first use \eqref{eq:1V1fsolseriesk0inf} to compute the sum up to \( m = 50 \), and give the function values at \((k_0,k_1)=(5,2)\):
\begin{equation}
\text{I}_{1}(5,2)=(2.81...  + i~ 1.68... ) \times 10^{-4} \, , ~~~\text{I}_{2}(5,2)=(3.38...  - i~ 5.65... )\times  10^{-4}
\end{equation}
We could choose the path with four steps:
\begin{equation}
(k_0,k_1)=(5,2)\to\left(\frac{7}{2}-i,2\right)\to(2-i,2)\to\left(\frac{3}{2}-\frac{i}{2},2\right)\to\left(\frac{3}{2},2\right)\,,
\end{equation}
where the final point satisfies \( k_0 < k_1 \) and is outside the convergence domain of \eqref{eq:1V1fsolseriesk0inf}. We compute them using \eqref{eq:seriesEqs} up to the 90th order at each point. Each step takes about 0.2 seconds. We obtained the final results with relative errors of \( \mathcal{O}(10^{-34}) \):
\begin{equation}
\text{I}_{1}(3/2,2)=0.201... + i~ 
 0.120...\, , ~~~\text{I}_{2}(3/2,2)=0.176...  - i~ 0.295... 
\end{equation}
Subsequently, one can use point \( (3/2,2) \) as a new boundary and solve for the function values in the region \( k_0 < k_1 \) along the real axis. Since the distance for each expansion is relatively short and the series solution converges quickly, we can obtain the values more efficiently. For instance, we tested to evaluate the master integrals at $(k_0,k_1)=(3/2-j/100,2)$, for $j=1,\cdots,100$ and with 20th order expansion. We find that each point took approximately 0.01 seconds to compute while maintaining a relative error of at most \( \mathcal{O}(10^{-34}) \). All these calculations were performed using Mathematica on a single-core CPU of a personal computer.
Additionally, one can use the function values at these regular points to match the expansion near the singularity \( k_1 \to \infty \) and determine the boundary condition coefficients for this expansion, as also has been discussed in the \cite{Moriello:2019yhu}.

\subsection{Example: 2-fold vertex integral family} \label{sec:1-v_2f}
\subsubsection{Preparation}

In this section, we use a 2-fold vertex integral family as an example to solve its series solution by the expansions of \( k_0 \to \infty \) and \( k_2 \to \infty \) using the $\nd \log$-form differential equations. This function family has 4 master integrals. Again, We use $\text{I}_{\tilde{\bm{a}}}$ as master integrals:
\begin{align}
&\text{I}_{1}=\int_{-\infty}^0 \nd \tau e^{ i k_0 \tau} (-\tau)^{\nu_0} h^{(2)}_{\nu_1}(0,-k_1\tau)h^{(2)}_{\nu_2}(0,-k_2\tau) \,,\nn\\ 
&\text{I}_{2}=\int_{-\infty}^0 \nd \tau e^{ i k_0 \tau} (-\tau)^{\nu_0} h^{(2)}_{\nu_1}(0,-k_1\tau) h^{(2)}_{\nu_2}(1,-k_2\tau)\,\nn\\
&\text{I}_{3}=\int_{-\infty}^0 \nd \tau e^{ i k_0 \tau} (-\tau)^{\nu_0} h^{(2)}_{\nu_1}(1,-k_1\tau)h^{(2)}_{\nu_2}(0,-k_2\tau) \,,\nn\\ 
&\text{I}_{4}=\int_{-\infty}^0 \nd \tau e^{ i k_0 \tau} (-\tau)^{\nu_0} h^{(2)}_{\nu_1}(1,-k_1\tau)h^{(2)}_{\nu_2}(1,-k_2\tau) \,.
\end{align}
The matrices of differential equations are
\begin{align}
\Omega&=A_1 \log \left(k_{--}\right)+A_2 \log \left(k_{-+}\right)+A_3 \log \left(k_{+-}\right)+A_4
   \log \left(k_{++}\right) \, \nn\\
   &~+A_5 \log \left(k_2\right)+A_6 \log \left(k_1\right) \,,
\end{align}
where
\begin{align}
&k_{\pm\pm}\equiv k_0\pm k_1\pm k_2 \, , \\
&A_1=\frac{1}{4}\left(
\begin{array}{cccc}
 -\nu _0-1 & i \left(\nu _0-2 \nu _2\right) & i \left(\nu _0-2 \nu _1\right) & \nu _0-2 \nu _1-2 \nu _2-1 \\
 -i \left(\nu _0+1\right) & 2 \nu _2-\nu _0 & 2 \nu _1-\nu _0 & i \left(\nu _0-2 \nu _1-2 \nu _2-1\right) \\
 -i \left(\nu _0+1\right) & 2 \nu _2-\nu _0 & 2 \nu _1-\nu _0 & i \left(\nu _0-2 \nu _1-2 \nu _2-1\right) \\
 \nu _0+1 & -i \left(\nu _0-2 \nu _2\right) & -i \left(\nu _0-2 \nu _1\right) & -\nu _0+2 \nu _1+2 \nu _2+1 \\
\end{array}
\right) \,,\nn\\
&A_2=\frac{1}{4}\left(
\begin{array}{cccc}
 -\nu _0-1 & -i \left(\nu _0-2 \nu _2\right) & i \left(\nu _0-2 \nu _1\right) & -\nu _0+2 \nu _1+2 \nu _2+1 \\
 i \left(\nu _0+1\right) & 2 \nu _2-\nu _0 & \nu _0-2 \nu _1 & i \left(\nu _0-2 \nu _1-2 \nu _2-1\right) \\
 -i \left(\nu _0+1\right) & \nu _0-2 \nu _2 & 2 \nu _1-\nu _0 & -i \left(\nu _0-2 \nu _1-2 \nu _2-1\right) \\
 -\nu _0-1 & -i \left(\nu _0-2 \nu _2\right) & i \left(\nu _0-2 \nu _1\right) & -\nu _0+2 \nu _1+2 \nu _2+1 \\
\end{array}
\right) \,,\nn\\
&A_3=\frac{1}{4}\left(
\begin{array}{cccc}
 -\nu _0-1 & i \left(\nu _0-2 \nu _2\right) & -i \left(\nu _0-2 \nu _1\right) & -\nu _0+2 \nu _1+2 \nu _2+1 \\
 -i \left(\nu _0+1\right) & 2 \nu _2-\nu _0 & \nu _0-2 \nu _1 & -i \left(\nu _0-2 \nu _1-2 \nu _2-1\right) \\
 i \left(\nu _0+1\right) & \nu _0-2 \nu _2 & 2 \nu _1-\nu _0 & i \left(\nu _0-2 \nu _1-2 \nu _2-1\right) \\
 -\nu _0-1 & i \left(\nu _0-2 \nu _2\right) & -i \left(\nu _0-2 \nu _1\right) & -\nu _0+2 \nu _1+2 \nu _2+1 \\
\end{array}
\right) \,,\nn\\
&A_4=\frac{1}{4}\left(
\begin{array}{cccc}
 -\nu _0-1 & -i \left(\nu _0-2 \nu _2\right) & -i \left(\nu _0-2 \nu _1\right) & \nu _0-2 \nu _1-2 \nu _2-1 \\
 i \left(\nu _0+1\right) & 2 \nu _2-\nu _0 & 2 \nu _1-\nu _0 & -i \left(\nu _0-2 \nu _1-2 \nu _2-1\right) \\
 i \left(\nu _0+1\right) & 2 \nu _2-\nu _0 & 2 \nu _1-\nu _0 & -i \left(\nu _0-2 \nu _1-2 \nu _2-1\right) \\
 \nu _0+1 & i \left(\nu _0-2 \nu _2\right) & i \left(\nu _0-2 \nu _1\right) & -\nu _0+2 \nu _1+2 \nu _2+1 \\
\end{array}
\right) \,,\nn\\
&A_5=\left(
\begin{array}{cccc}
 0 & 0 & 0 & 0 \\
 0 & 0 & 0 & 0 \\
 0 & 0 & -2 \nu _1-1 & 0 \\
 0 & 0 & 0 & -2 \nu _1-1 \\
\end{array}
\right) \,,~~~
A_6\left(
\begin{array}{cccc}
 0 & 0 & 0 & 0 \\
 0 & -2 \nu _2-1 & 0 & 0 \\
 0 & 0 & 0 & 0 \\
 0 & 0 & 0 & -2 \nu _2-1 \\
\end{array}
\right)  \, .
\end{align}

\subsubsection{Solutions with the boundary $k_0\to \infty$} 

Defining $x=1/k_0$ and following the same steps, indicial equations give
\ie
&\text{Solution 1: }\text{C}(i\neq 1,0)= 0,~~~\lambda= \nu_0+1 \, ,\\
&\text{Solution 2: }\text{C}(i\neq 2,0)= 0,~~~\lambda= \nu_0-2\nu_2 \, ,\\
&\text{Solution 3: }\text{C}(i\neq 3,0)= 0,~~~\lambda= \nu_0-2 \nu_1 \, ,\\
&\text{Solution 4: }\text{C}(i\neq 4,0)= 0,~~~\lambda= \nu_0-2 \nu_1-2\nu_2-1 \, .
\fe
We have four solutions:
\begin{align}
&\text{I}_{\tilde{\bm{a}}}= \sum_{\tilde{\bm{b}}=1}^4 \text{C}^{[\tilde{\bm{b}}]}f_{\tilde{\bm{a}}}^{[\tilde{\bm{b}}]}\, , \nn\\
&f_{1}^{[1]}=x^{\nu _0+1} \text{F}_4\left(\frac{\nu _0+1}{2} ,\frac{\nu _0+2}{2};\nu _1+1,\nu
   _2+1;k_1^2 x^2,k_2^2 x^2\right) \, , \nn\\
&f_{2}^{[1]}=x^{\nu _0+2}\frac{i k_2 \left(\nu _0+1\right) }{2 \left(\nu _2+1\right)} 
\text{F}_4\left(\frac{\nu_0+2}{2},\frac{\nu _0+3}{2} ;\nu_1+1,\nu_2+2;k_1^2 x^2,k_2^2 x^2\right)\, , \nn\\
&f_{3}^{[1]}= x^{\nu _0+2}\frac{i k_1 \left(\nu _0+1\right)}{2 \left(\nu _1+1\right)}
\text{F}_4\left(\frac{\nu_0+2}{2},\frac{\nu_0+3}{2} ;\nu_1+2,\nu_2+1;k_1^2 x^2,k_2^2 x^2\right)  \, , \nn\\
&f_{4}^{[1]}=x^{\nu _0+3}\frac{-k_1 k_2 \left(\nu _0+1\right) \left(\nu _0+2\right) }{4 \left(\nu _1+1\right)\left(\nu _2+1\right)}
\text{F}_4\left(\frac{\nu _0+3}{2},\frac{\nu _0+4}{2};\nu_1+2,\nu_2+2;k_1^2 x^2,k_2^2 x^2\right)  \, , \nn\\
&f_{1}^{[2]}=x^{\nu _0-2 \nu _2+1}\frac{i k_2 \left(\nu _0-2 \nu_2\right) }{2 \nu_2}
\text{F}_4\left(\frac{\nu_0-2 \nu_2+1}{2},\frac{\nu_0-2\nu_2+2}{2};\nu_1+1,1-\nu_2;k_1^2 x^2,k_2^2 x^2\right) \, , \nn\\
&f_{2}^{[2]}=x^{\nu_0-2\nu_2}  
\text{F}_4\left(\frac{\nu_0-2 \nu_2}{2},\frac{\nu _0-2 \nu_2+1}{2} ;\nu _1+1,-\nu _2;k_1^2 x^2,k_2^2 x^2\right)  \, , \nn\\
&f_{3}^{[2]}= x^{\nu _0-2 \nu _2+2}\frac{ - k_1 k_2 \left(\nu _0-2 \nu _2\right) \left(\nu _0-2 \nu _2+1\right)}{4 \nu_2 \left(\nu _1+1\right)} \nn\\
&~~~~~~~~\times \text{F}_4\left(\frac{\nu_0-2 \nu_2+2}{2},\frac{\nu _0-2 \nu_2+3}{2} ;\nu _1+2,1-\nu _2;k_1^2 x^2,k_2^2 x^2\right)\, , \nn\\
&f_{4}^{[2]}=x^{\nu _0-2 \nu _2+1}\frac{i k_1 \left(\nu _0-2 \nu _2\right) }{2 \left(\nu_1+1\right)}
\text{F}_4\left(\frac{\nu_0-2 \nu_2+1}{2} ,\frac{\nu _0-2 \nu_2+2}{2} ;\nu _1+2,-\nu _2;k_1^2 x^2,k_2^2 x^2\right)  \, , \nn\\
&f_{1}^{[3]}=x^{\nu _0-2 \nu _1+1}\frac{i k_1 \left(\nu _0-2 \nu _1\right) }{2 \nu _1}
\text{F}_4\left(\frac{\nu_0-2 \nu _1+1}{2},\frac{\nu _0-2 \nu _1+2}{2};1-\nu_1,\nu_2+1;k_1^2 x^2,k_2^2 x^2\right) \, , \nn\\
&f_{2}^{[3]}=x^{\nu _0-2 \nu _1+2} \frac{-k_1 k_2 \left(\nu _0-2 \nu _1\right) \left(\nu _0-2 \nu _1+1\right) }{4 \nu_1 \left(\nu _2+1\right)} \nn\\
&~~~~~~~~\times \text{F}_4\left(\frac{\nu_0-2 \nu_1+2}{2},\frac{\nu _0-2 \nu_1+3}{2} ;1-\nu _1,\nu _2+2;k_1^2 x^2,k_2^2 x^2\right)\, , \nn\\
&f_{3}^{[3]}=x^{\nu_0-2\nu_1}  
\text{F}_4\left(\frac{\nu_0-2 \nu_1}{2},\frac{\nu _0-2 \nu_1+1}{2} ;-\nu _1,\nu _2+1;k_1^2 x^2,k_2^2 x^2\right)  \, , \nn\\
&f_{4}^{[3]}=x^{\nu _0-2 \nu _1+1}\frac{i k_2 \left(\nu _0-2 \nu _1\right) }{2 \left(\nu_2+1\right)}
\text{F}_4\left(\frac{\nu_0-2 \nu_1+1}{2},\frac{\nu _0-2 \nu_1+2}{2} ;-\nu _1,\nu _2+2;k_1^2 x^2,k_2^2 x^2\right)  \, , \nn\\
&f_{1}^{[4]}=x^{\nu _0-2 \nu _1-2 \nu _2+1}\frac{-k_1 k_2 \left(\nu _0-2 \nu _1-2 \nu _2-1\right) \left(\nu _0-2 \nu _1-2\nu _2\right)
   }{4 \nu _1 \nu _2} \nn\\
&~~~~~~~~\times \text{F}_4\left(\frac{1}{2} \left(\nu _0-2\nu _1-2\nu_2+1\right),\frac{1}{2} \left(\nu _0-2\nu _1-2\nu _2+2\right);1-\nu _1,1-\nu _2;k_1^2
   x^2,k_2^2 x^2\right)  \, , \nn\\
&f_{2}^{[4]}= x^{\nu _0-2 \nu _1-2 \nu _2}\frac{i k_1 \left(\nu _0-2 \nu _1-2 \nu _2-1\right)}{2 \nu
   _1} \nn\\
&~~~~~~~~\times \text{F}_4\left(\frac{1}{2} \left(\nu _0-2\nu _1-2\nu _2\right),\frac{1}{2} \left(\nu _0-2\nu _1-2\nu _2+1\right);1-\nu _1,-\nu _2;k_1^2 x^2,k_2^2 x^2\right)  \, , \nn\\
&f_{3}^{[4]}= x^{\nu _0-2 \nu _1-2 \nu _2}\frac{i k_2 \left(\nu _0-2 \nu _1-2 \nu _2-1\right) }{2 \nu_2} \nn\\
&~~~~~~~~\times \text{F}_4\left(\frac{1}{2} \left(\nu _0-2\nu _1-2\nu _2\right),\frac{1}{2} \left(\nu _0-2\nu _1-2\nu _2+1\right);-\nu _1,1-\nu _2;k_1^2 x^2,k_2^2 x^2\right)  \, , \nn\\
&f_{4}^{[4]}= x^{\nu _0-2 \nu _1-2 \nu _2-1} \nn\\
&~~~~~~~~\times \text{F}_4\left(\frac{1}{2} \left(\nu _0-2 \nu _1 -2\nu _2-1\right),\frac{1}{2} \left(\nu _0-2 \nu _1-2 \nu _2\right);-\nu _1,-\nu _2;k_1^2
   x^2,k_2^2 x^2\right)  \, .
\end{align}
Here the $\text{F}_4$ are \cite{Bera:2023pyz}
\begin{equation}
\text{F}_4(a,b;c_1,c_2;x,y)\equiv \sum_{m,n=0}^\infty \frac{(a)_{m+n}(b)_{m+n}}{(c_1)_m(c_2)_n}  \frac{x^m y^n}{m! n!} \,.
\end{equation}

To determine $\text{C}^{[\tilde{\bm{b}}]}$, we only need one term in the expansion of integrand again. For example, referring to \eqref{eq:expandha2R}, we have
\begin{align}
&\text{I}_{1}\sim\int_{-\infty}^0 \nd \tau e^{ i k_0 \tau} (-\tau)^{\nu_0}
\left(\text{C}_1^{(k_0)}(\nu_1)  +  \mathcal{O}(t^{-2\nu_1})\right)
\left(\text{C}_1^{(k_0)}(\nu_2) +  \mathcal{O}(\tau^{-2\nu_2}) \right) \,.
\end{align}
Here only the $\text{C}_1^{(k_0)}(\nu_1)\text{C}_1^{(k_0)}(\nu_2)$ contributes to leading-order of solution one, and thus it contributes to $\text{C}^{[1]}$. Meanwhile, $\text{C}_1^{(k_0)}(\nu_1)\mathcal{O}(\tau^{-2\nu_2})$ contributes to the second and high order of solution two, $\text{C}_1^{(k_0)}(\nu_2)\mathcal{O}(\tau^{-2\nu_1})$ contributes to the second and high order of solution three, $\mathcal{O}(\tau^{-2\nu_2})\mathcal{O}(\tau^{-2\nu_1})$ contributes to the second and high order of solution four. Through similar analysis, we focus only on the following terms in the expansion (recalling the definition \eqref{eq:defindex} that $\text{I}_{\{a_1,a_2\}}\equiv\text{I}_{\tilde{\bm{a}}}$):
\begin{align}
&\text{I}_{\{0,0\}}\sim \text{C}^{[\{0,0\}]} x^{\nu_0+1} = \int_{-\infty}^0 \nd \tau ~ \text{C}_{\tilde{0}}^{(k_0)}(\nu_1)~ \text{C}_{\tilde{0}}^{(k_0)}(\nu_2) ~ e^{ i k_0 \tau} (-\tau)^{\nu_0} \,, \nn\\
&\text{I}_{\{0,1\}}\sim \text{C}^{[\{0,1\}]} x^{\nu_0-2\nu_2} = \int_{-\infty}^0 \nd \tau ~ \text{C}_{\tilde{0}}^{(k_0)}(\nu_1) ~ \text{C}_{\tilde{1}}^{(k_0)}(\nu_2) ~ e^{ i k_0 \tau} (-\tau)^{\nu_0}(-k_2\tau)^{-2\nu_2-1} \,, \nn\\
&\text{I}_{\{1,0\}}\sim \text{C}^{[\{1,0\}]} x^{\nu_0-2\nu_2} = \int_{-\infty}^0 \nd \tau ~ \text{C}_{\tilde{1}}^{(k_0)}(\nu_1) ~ \text{C}_{\tilde{0}}^{(k_0)}(\nu_2) ~ e^{ i k_0 \tau} (-\tau)^{\nu_0}(-k_1\tau)^{-2\nu_1-1} \,, \nn\\
&\text{I}_{\{1,1\}}\sim \text{C}^{[\{1,1\}]} x^{\nu_0-2\nu_2-2\nu_1-1} \nn\\
&~~~~~~~= \int_{-\infty}^0 \nd \tau ~ \text{C}_{\tilde{1}}^{(k_0)}(\nu_1) ~ \text{C}_{\tilde{1}}^{(k_0)}(\nu_2) ~ e^{ i k_0 \tau} (-\tau)^{\nu_0}(-k_1\tau)^{-2\nu_1-1}(-k_2\tau)^{-2\nu_2-1} \,. 
\end{align}
Hence,
\begin{align}
&\text{C}^{[\{0,0\}]}=(-i)^{\nu _0+1} \Gamma \left(\nu _0+1\right) ~ \text{C}_{\tilde{0}}^{(k_0)}(\nu_1)~ \text{C}_{\tilde{0}}^{(k_0)}(\nu_2) \, , \nn\\
&\text{C}^{[\{0,1\}]}=(-i)^{\nu _0-2 \nu _2} k_2^{-2 \nu _2-1} \Gamma \left(\nu _0-2 \nu _2\right) ~ \text{C}_{\tilde{0}}^{(k_0)}(\nu_1)~ \text{C}_{\tilde{1}}^{(k_0)}(\nu_2) \, , \nn\\
&\text{C}^{[\{0,1\}]}=(-i)^{\nu _0-2 \nu _1} k_1^{-2 \nu _1-1} \Gamma \left(\nu _0-2 \nu _1\right) ~ \text{C}_{\tilde{1}}^{(k_0)}(\nu_1)~ \text{C}_{\tilde{0}}^{(k_0)}(\nu_2) \, ,\nn\\
&\text{C}^{[\{1,1\}]}= (-i)^{\nu _0-2 \left(\nu _1+\nu _2\right)-1} k_1^{-2 \nu _1-1} k_2^{-2 \nu _2-1} \Gamma \left(\nu _0-2 \nu _1-2\nu _2-1\right) ~ \text{C}_{\tilde{1}}^{(k_0)}(\nu_1)~ \text{C}_{\tilde{1}}^{(k_0)}(\nu_2) \, .
\end{align}

\subsubsection{Solutions with the boundary $k_2\to\infty$}

Without loss of generality, we consider the expansion near the boundary as $k_2\to\infty$.
Defining $x=1/k_2$ and following the same steps, indicial equations give
\begin{align}
&\text{Solution 1 \& 2: }\lambda= \nu_0+1 \, , ~~~~ \text{C}(3,0)=\text{C}(4,0)= 0  \,, \nn\\
&\text{Solution 3 \& 4: }\lambda= \nu_0-2 \nu_1  \, , ~~ \text{C}(1,0)=\text{C}(2,0)= 0  \, . 
\end{align}
We denote $\text{C}(1,0)$ and $\text{C}(2,0)$ in solution $1\&2$ by $\text{C}^{[1]}$ and $\text{C}^{[2]}$, denote $\text{C}(3,0)$ and $\text{C}(4,0)$ in solution $3\&4$ by $\text{C}^{[3]}$ and $\text{C}^{[4]}$, and have four solutions:
\begin{align}
&\text{I}_{\tilde{\bm{a}}}= \sum_{\tilde{\bm{b}}=1}^4 \text{C}^{[\tilde{\bm{b}}]}f_{\tilde{\bm{a}}}^{[\tilde{\bm{b}}]}\, , \nn\\
&f_{1}^{[1]}=x^{\nu _0+1}
\text{F}_4\left(\frac{1}{2} \left(\nu _0+1\right),\frac{1}{2} \left(\nu _0-2 \nu _2+1\right);\frac{1}{2},\nu _1+1;k_0^2 x^2,k_1^2 x^2\right) \, , \nn\\ 
&f_{2}^{[1]}= x^{\nu _0+2} i k_0 \left(\nu _0+1\right)
\text{F}_4\left(\frac{1}{2} \left(\nu _0+3\right) , \frac{1}{2} \left(\nu _0-2 \nu _2+1\right);\frac{3}{2},\nu _1+1 ; k_0^2 x^2,k_1^2 x^2\right) \, , \nn\\ 
&f_{3}^{[1]}=x^{\nu _0+3} \frac{-i k_0 k_1 \left(\nu _0+1\right) \left(\nu _0-2 \nu _2+1\right) }{2 \left(\nu _1+1\right)} \nn\\
&~~~~~~~\times
\text{F}_4\left(\frac{1}{2} \left(\nu _0+3\right),\frac{1}{2} \left(\nu _0-2 \nu _2+3\right);\frac{3}{2},\nu _1+2;k_0^2 x^2,k_1^2 x^2\right)     \, , \nn\\ 
&f_{4}^{[1]}= x^{\nu _0+2} \frac{k_1 \left(\nu _0+1\right) }{2 \left(\nu _1+1\right)}
\text{F}_4\left(\frac{1}{2} \left(\nu _0+3\right),\frac{1}{2} \left(\nu _0-2 \nu _2+1\right);\frac{1}{2},\nu _1+2;k_0^2 x^2,k_1^2 x^2\right)   
   \, , \nn\\ 
&f_{1}^{[2]}=
x^{\nu _0+2} (-i k_0) \left(\nu _0-2 \nu _2\right) 
\text{F}_4\left(\frac{1}{2} \left(\nu _0+2\right),\frac{1}{2} \left(\nu _0-2 \nu _2+2\right);\frac{3}{2},\nu _1+1;k_0^2 x^2,k_1^2 x^2\right)      \, , \nn\\ 
&f_{2}^{[2]}=
x^{\nu _0+1}
\text{F}_4\left(\frac{1}{2} \left(\nu _0+2\right),\frac{1}{2} \left(\nu _0-2 \nu _2\right);\frac{1}{2},\nu _1+1;k_0^2 x^2,k_1^2 x^2\right)       \, , \nn\\ 
&f_{3}^{[2]}=
x^{\nu _0+2} \frac{- k_1 \left(\nu _0-2 \nu _2\right) }{2 \left(\nu _1+1\right)}
\text{F}_4\left(\frac{1}{2} \left(\nu _0+2\right),\frac{1}{2} \left(\nu _0-2 \nu _2+2\right);\frac{1}{2},\nu _1+2;k_0^2 x^2,k_1^2 x^2\right)       \, , \nn\\ 
&f_{4}^{[2]}=
 x^{\nu _0+3} \frac{-i k_0 k_1 \left(\nu _0+2\right) \left(\nu _0-2 \nu _2\right)}{2 \left(\nu _1+1\right)}   
 \nn\\
&~~~~~~~\times
\text{F}_4\left(\frac{1}{2} \left(\nu _0+4\right),\frac{1}{2} \left(\nu _0-2 \nu _2+2\right);\frac{3}{2},\nu _1+2;k_0^2 x^2,k_1^2 x^2\right)  
      \, , \nn\\ 
&f_{1}^{[3]}=
x^{\nu _0-2 \nu _1+2}
\frac{-i k_0 k_1 \left(\nu _0-2 \nu _1\right) \left(\nu _0-2 \left(\nu _1+\nu _2\right)\right) }{2 \nu _1} \nn\\
&~~~~~~~\times
\text{F}_4\left(\frac{1}{2} \left(\nu _0-2 \left(\nu _1+\nu _2-1\right)\right),\frac{1}{2} \left(\nu _0-2 \nu _1+2\right);\frac{3}{2},1-\nu _1;k_0^2 x^2,k_1^2 x^2\right)      \, , \nn\\ 
&f_{2}^{[3]}=
x^{\nu _0-2 \nu _1+1}
\frac{k_1 \left(\nu _0-2 \nu _1\right) }{2 \nu _1}
\text{F}_4\left(\frac{1}{2} \left(\nu _0-2 \left(\nu _1+\nu _2\right)\right),\frac{1}{2} \left(\nu _0-2 \nu _1+2\right);\frac{1}{2},1-\nu _1;k_0^2 x^2,k_1^2 x^2\right)       \, , \nn\\ 
&f_{3}^{[3]}=
x^{\nu _0-2 \nu _1}
\text{F}_4\left(\frac{1}{2} \left(\nu _0-2 \left(\nu _1+\nu _2\right)\right),\frac{1}{2} \left(\nu _0-2 \nu _1\right);\frac{1}{2},-\nu _1;k_0^2 x^2,k_1^2 x^2\right)       \, , \nn\\ 
&f_{4}^{[3]}=
x^{\nu _0-2 \nu _1+1}
i k_0 \left(\nu _0-2 \nu _1\right) 
\text{F}_4\left(\frac{1}{2} \left(\nu _0-2 \left(\nu _1+\nu _2\right)\right),\frac{1}{2} \left(\nu _0-2 \nu _1+2\right);\frac{3}{2},-\nu _1;k_0^2 x^2,k_1^2 x^2\right)  
      \, , \nn\\ 
&f_{1}^{[4]}=
x^{\nu _0-2 \nu _1+1}
\frac{k_1 \left(-\nu _0+2 \nu _1+2 \nu _2+1\right) }{2 \nu _1}   \nn\\
&~~~~~~~\times
\text{F}_4\left(\frac{1}{2} \left(\nu _0-2 \nu _1-2 \nu _2+1\right),\frac{1}{2} \left(\nu _0-2 \nu _1+1\right);\frac{1}{2},1-\nu _1;k_0^2 x^2,k_1^2 x^2\right)      \, , \nn\\ 
&f_{2}^{[4]}=
x^{\nu _0-2 \nu _1+2}\frac{-i k_0 k_1 \left(\nu _0-2 \nu _1+1\right) \left(\nu _0-2 \nu _1-2 \nu _2-1\right) }{2 \nu _1}   \nn\\
&~~~~~~~\times
\text{F}_4\left(\frac{1}{2} \left(\nu _0-2 \nu _1-2 \nu _2+1\right),\frac{1}{2} \left(\nu _0-2 \nu _1+3\right);\frac{3}{2},1-\nu _1;k_0^2 x^2,k_1^2 x^2\right)       \, , \nn\\ 
&f_{3}^{[4]}=
x^{\nu _0-2 \nu _1+1} (-i k_0) \left(\nu _0-2 \nu _1-2 \nu _2-1\right)   \nn\\
&~~~~~~~\times 
\text{F}_4\left(\frac{1}{2} \left(\nu _0-2 \nu _1-2 \nu _2+1\right),\frac{1}{2} \left(\nu _0-2 \nu _1+1\right);\frac{3}{2},-\nu _1;k_0^2 x^2,k_1^2 x^2\right)       \, , \nn\\ 
&f_{4}^{[4]}=
x^{\nu _0-2 \nu _1}
\text{F}_4\left(\frac{1}{2} \left(\nu _0-2 \nu _1-2 \nu _2-1\right),\frac{1}{2} \left(\nu _0-2 \nu _1+1\right);\frac{1}{2},-\nu _1;k_0^2 x^2,k_1^2 x^2\right)  
      \, .
\end{align}

Recalling \eqref{eq:Hasyinf}, all terms in the integrand, except for the Hankel function corresponding to $k_2$, are exponentially suppressed and thus can be expanded around $\tau=0$. Then, recalling \eqref{eq:Hasy0}, all boundary coefficients could be determined by integrals taking the form just like the two in \eqref{eq:1V1fBCCk1inf}. Let us consider $\text{I}_{3}$ as an example. In the $k_2\to \infty$ limitation,
\begin{align}
&\text{I}_{3}=\int_{-\infty}^0 \nd \tau e^{ i k_0 \tau} (-\tau)^{\nu_0} h^{(2)}_{\nu_1}(1,-k_1\tau) h^{(2)}_{\nu_2}(0,-k_2\tau) \nn\\
&~~~ \sim \int_{-\infty}^0 \nd \tau e^{ i k_0 \tau} (-\tau)^{\nu_0} (-k_1\tau)^{} \text{C}_2^{(k_0)}(\nu_1) h^{(2)}_{\nu_2}(1,-k_2\tau) \nn\\
&~~~ = \text{C}_2^{(k_0)}(\nu_1)  \text{C}_1^{(k_n)}(\nu_0-2\nu_1-1,\nu_2) ~ k_1^{-2\nu_1-1}  x^{\nu_0-2\nu_1} \nn\\
&~~~ = \text{C}^{[3]}  x^{\nu_0-2\nu_1} \,.
\end{align}
As a result, we have 
\begin{align}
&\text{C}^{[1]}=\text{C}^{(k_0)}_1\left(\nu _1\right) \text{C}^{(k_n)}_1\left(\nu _0,\nu _2\right)   \,, \nn\\
&\text{C}^{[2]}=\text{C}^{(k_0)}_1\left(\nu _1\right) \text{C}^{(k_n)}_2\left(\nu _0,\nu _2\right) \,, \nn\\
&\text{C}^{[3]}= \text{C}^{(k_0)}_2\left(\nu _1\right) \text{C}^{(k_n)}_1\left(\nu _0-2 \nu _1-1,\nu _2\right) k_1^{-2 \nu _1-1} \,, \nn\\
&\text{C}^{[4]}=  \text{C}^{(k_0)}_2\left(\nu _1\right) \text{C}^{(k_n)}_2\left(\nu _0-2 \nu _1-1,\nu _2\right) k_1^{-2 \nu _1-1} \,.
\end{align}

\subsection{Multivariate hypergeometric solutions of arbitrary vertex integral family}   \label{sec:1-v_nf}
\subsubsection{Solutions with the boundary $k_0\to\infty$}

For the master integrals of an arbitrary $n$-fold vertex function family with  $h^{(2)}_\nu(a,-k\tau)$ and all  $\nu_i$ being real (we will generalize the results to general cases later in Sec \ref{sec:generalC}), we could easily derive $\text{C}^{[\tilde{\bm{b}}]}$ by computation similar to previous subsections, and by observation, we conjecture  solutions $f_{\tilde{\bm{a}}}^{[\tilde{\bm{b}}]}$  of expansion $k_0\to\infty$ as follows:
\begin{align}
&\text{I}_{\tilde{\bm{a}}}= \sum_{\tilde{\bm{b}}=1}^{2^n} \text{C}^{[\tilde{\bm{b}}]}f_{\tilde{\bm{a}}}^{[\tilde{\bm{b}}]}\, , \nn\\
&f_{\tilde{\bm{a}}}^{[\tilde{\bm{b}}]}=x^{\tilde{\text{A}}} 
\frac{(\tilde{\text{A}})_{|\bm{a}-\bm{b}|\cdot \bm{1}} }{2^{|\bm{a}-\bm{b}|\cdot \bm{1}}} 
\prod_{j=1}^n \left(\frac{(-1)^{b_j}  
 i k_j x }{\tilde{\text{B}}_j}\right)^{|a_j-b_j|} \nn\\
 &~~~~~~~~\times 
\tilde{\text{F}}_4\left(\text{A}_1,\text{A}_2;~\text{B}_1,\cdots,\text{B}_n;~k_1^2x^2,\cdots,k_n^2x^2\right) 
 \,  , \nn\\
 &\tilde{\text{A}}=\nu_0+1-\bm{b}\cdot (        2\bm{\nu}+\bm{1})\,,~~~~~~~~~~ \text{A}_j = \frac{1}{2}\left(\tilde{\text{A}}+|\bm{a}-\bm{b}|\cdot \bm{1}-1 + j\right)  \, ,\nn\\
 & \tilde{\text{B}}_j = \nu_j+1-b_j(2\nu_j+1) \, , ~~~~~~~~~\text{B}_j= \tilde{\text{B}}_j+|a_j-b_j|  \,, \\
 &\text{C}^{[\tilde{\bm{b}}]} = (-i)^{\nu_0+1} \Gamma(\tilde{\text{A}}) \prod_{j=1}^n (-ik_i)^{-b_j(2\nu_j+1)}\text{C}_{\tilde{b}_j}^{(k_0)}(\nu_j) \, ,
\end{align}
where
\begin{align}
&\tilde{\text{F}}_4\left(\text{A}_1,\text{A}_2;\textbf{B};\bm{z}\right)\equiv \sum_{m_1,\cdots,m_n=0}^\infty \frac{(\text{A}_1)_{\bm{m}\cdot \bm{1}}(\text{A}_2)_{\bm{m}\cdot \bm{1}}}{\prod_{i=1}^n(\text{B}_i)_{m_i}}  \prod_{i=1}^n \frac{z_i^{m_i}}{m_i!}  \, , \nn\\
&\bm{a}=a_1,a_2,\cdots,a_n \,~,~~~~\bm{b}=b_1,b_2,\cdots,b_n \, ~, \nn\\
&\bm{\nu}=\nu_1,\nu_2,\cdots,\nu_n  \,~,~~~~\bm{1}=1,1,\cdots,1 \, ~,\nn\\
&|\bm{a}-\bm{b}|=|a_1-b_1|,|a_2-b_2|,\cdots,|a_n-b_n| \,~.
\end{align}
We have verified the result to be right by comparing it with both the power series solution from differential equations and results of direct numerical integration, up to $n=4$.

\subsubsection{Solutions with the boundary $k_n\to\infty$}

Similarly, the multivariate hypergeometric solutions around $k_n\to \infty$ also could be given by
\begin{align}
&\text{I}_{\tilde{\bm{a}}}= \sum_{\tilde{\bm{b}}=1}^{2^n} \text{C}^{[\tilde{\bm{b}}]}f_{\tilde{\bm{a}}}^{[\tilde{\bm{b}}]}\, , \nn\\
&f_{\tilde{\bm{a}}}^{[\tilde{\bm{b}}]}=
x^{\nu_0+1-\hat{\bm{b}}\cdot (2\hat{\bm{\nu}}+\bm{1})}  (-1)^{\lfloor \left( \text{mod}[|\bm{a}-\bm{b}|\cdot \bm{1},2]+\hat{\bm{b}}\cdot \bm{1} +\text{mod}[\tilde{\bm{b}} - 1, 2]\right) / 2 \rfloor}  \nn\\
 &~~~~~~~~\times  (2 i k_0 x)^{\text{mod}[|\bm{a}-\bm{b}|\cdot \bm{1},2]} \prod_{j=1}^{n-1}
 \left( \frac{(-1)^{b_j} k_j x}{\nu_j+1-b_j(2\nu_j+1)} \right)^{|a_j-b_j|}     \nn\\
 &~~~~~~~~\times \prod_{j=0}^{\text{mod}[|\bm{a}-\bm{b}|\cdot \bm{1},2]+\hat{\bm{b}}\cdot \bm{1}-1} \left(       \frac{\nu_j+1-b_j(2\nu_j+1)}{2} - \text{mod}[b_n+j,2]\frac{2\nu_n+1}{2}  \right)      \nn\\
&~~~~~~~~\times\tilde{\text{F}}_4\left(\text{A}_1,\text{A}_2;~\text{B}_0,\text{B}_1,\cdots,\text{B}_{n-1};~k_0^2x^2,\cdots,k_{n-1}^2x^2\right) 
 \,  ,
\end{align}
where
 \begin{align}
 & \text{A}_i = \left(  \frac{\nu_j+1-b_j(2\nu_j+1)}{2} - \text{mod}[b_n+j,2]\frac{2\nu_n+1}{2}  \right)\Big|_{j=\text{mod}[|\bm{a}-\bm{b}|\cdot \bm{1},2]+\hat{\bm{b}}\cdot \bm{1}-1+i}       \, ,\nn\\
 & \text{B}_0 =\frac{1}{2}+\text{mod}[|\bm{a}-\bm{b}|\cdot \bm{1},2]        \, ,\nn\\
 &\text{B}_{i>0}=\nu_i+1-b_i(2\nu_i+1) +|a_i-b_i|        \,, \\
 &\text{C}^{[\tilde{\bm{b}}]} = \text{C}_{\tilde{b}_n}^{(k_n)}(\nu_0-\hat{\bm{b}}\cdot (2\hat{\bm{\nu}}+\bm{1}),\nu_n) \prod_{j=1}^{n-1} k_j^{-b_j(2\nu_j+1)} \text{C}_{\tilde{b}_j}^{(k_0)}(\nu_j) \, ,
\end{align}
$\text{mod}[a,b]$ represents the remainder when a is divided by b, the  $\lfloor ~  \rfloor$  indicates rounding up, and
\begin{align}
&\hat{\bm{a}}=a_1,a_2,\cdots,a_{n-1} \, ~,~~~~\hat{\bm{b}}=b_1,b_2,\cdots,b_{n-1} \, ~ ,~~~~  \hat{\bm{\nu}}=\nu_1,\nu_2,\cdots,\nu_{n-1} \, .
\end{align}
We have verified the result to be right by comparing it with power series solutions, which are directly solved from differential equations, up to $n=5$ and $\mathcal{O}[x^{\lambda+20}]$.

\subsubsection{Results for $h^{(1,2)}$ and real/imaginary $\nu$} \label{sec:generalC}

Now, let us discuss more general cases: each $h$-function in $n$-fold Hankel vertex integral family could  be  $h^{(1)}_\nu(a,-k\tau)$ or $h^{(2)}_\nu(a,-k\tau)$, and each $\nu_i$ could be real or  imaginary.
Since for all these cases, the differential equations are the same \cite{Chen:2023iix}, we have the same solutions $f_{\tilde{\bm{a}}}^{[\tilde{\bm{b}}]}$. However, the  $\text{C}_{\tilde{a}}^{(k_0)}(\nu)$ and $\text{C}_{\tilde{a}_n}^{(k_n)}(\nu,\nu_n)$ in the boundary coefficients $\text{C}^{[\tilde{\bm{b}}]}$ could be changed and should be re-determined. $\text{C}_{\tilde{a}}^{(k_0)}(\nu)$ should be re-determined by \eqref{eq:asyhRInu} like \eqref{eq:expandha2R}. For ease of reading, we  recall \eqref{eq:Hasy0} \eqref{eq:asyhRInu} and  \eqref{eq:expandha2R} here:
\begin{align}
&h_\nu^{(1)}(0,-k\tau)=c_1 (1+\mathcal{O}(\tau^2)) +  c_2 (-k\tau)^{-2\nu} (1+\mathcal{O}(\tau^2)) \, , \nn\\
&h_\nu^{(2)}(0,-k\tau)=c_1^{\star} (-k\tau)^{-\nu+\nu^\star}(1+\mathcal{O}(\tau^2)) +  c_2^{\star} (-k\tau)^{-\nu-\nu^\star} (1+\mathcal{O}(\tau^2)) \, ,\nn\\
&c_1=e^{-i\pi\nu}c[\nu]\,, ~~ c_2=c[-\nu]\,, ~~
c[\nu]\equiv \frac{ 2^{-\nu } \Gamma (-\nu )}{i \pi } \,.
\end{align}
For  $h^{(2)}_\nu(a,-k\tau)$ with real $\nu$, the corresponding $\text{C}_1^{(k_0)}(\nu)$ in the boundary coefficients have been determined as follow:
\begin{align}
&h_{\nu}^{(2)}(0,-k\tau)\sim c_1^\star +  c_2^\star (-k\tau)^{-2\nu}  = \text{C}_1^{(k_0)}(\nu)  +  \mathcal{O}(\tau^{-2\nu})   \, , \nn\\
&h_{\nu}^{(2)}(1,-k\tau)\sim-2\nu c_2^\star (-k\tau)^{-2\nu-1}=\text{C}_2^{(k_0)}(\nu) (-k\tau)^{-2\nu-1}\, ,\nn\\
&\text{C}_1^{(k_0)}(\nu) = c_1^\star=-e^{i\pi\nu}c[\nu] \, , ~~\text{C}_2^{(k_0)}(\nu) = -2\nu c_2^\star=c[-\nu-1] \, ;  \label{eq:expandha2Iagain}
\end{align}
For a $h^{(2)}_{\nu}(a,-k\tau)$ with  imaginary  $\nu$, the corresponding $\text{C}_1^{(k_0)}(\nu)$ in the boundary coefficients are determined as follow:
\begin{align}
&h_{\nu}^{(2)}(0,-k\tau)\sim c_2^\star +  c_1^\star (-k\tau)^{-2\nu}  = \text{C}_1^{(k_0)}(\nu)  +  \mathcal{O}(\tau^{-2\nu})   \, , \nn\\
&h_{\nu}^{(2)}(1,-k\tau)\sim-2\nu c_1^\star (-k\tau)^{-2\nu-1}=\text{C}_2^{(k_0)}(\nu) (-k\tau)^{-2\nu-1}\, ,\nn\\
&\text{C}_1^{(k_0)}(\nu) = c_2^\star=-c[\nu] \, , ~~\text{C}_2^{(k_0)}(\nu) = -2\nu c_1^\star=e^{-i\pi \nu}c[-\nu-1] \, ; \label{eq:expandha2I}
\end{align}
Since for both cases that   $\nu$ is  real or  imaginary in  $h^{(1)}_{\nu}(a,-k\tau)$, they do not involve $\nu^\star$ at the beginning, these two cases have the same corresponding $\text{C}_{\tilde{a}}^{(k_0)}(\nu)$:
\begin{align}
&h_{\nu}^{(1)}(0,-k\tau)\sim c_1 +  c_2 (-k\tau)^{-2\nu}  = \text{C}_1^{(k_0)}(\nu)  +  \mathcal{O}(\tau^{-2\nu})   \, , \nn\\
&h_{\nu}^{({1})}(1,-k\tau)\sim -2\nu c_2 (-k\tau)^{-2\nu-1}=\text{C}_2^{(k_0)}(\nu) (-k\tau)^{-2\nu-1}\, ,\nn\\
&\text{C}_1^{(k_0)}(\nu) = c_1=e^{-i\pi\nu}c[\nu] \, , ~~\text{C}_2^{(k_0)}(\nu) = -2\nu c_2=-c[-\nu-1] \, , \label{eq:expandha1}
\end{align}

Let's turn to consider $\text{C}_{\tilde{a}_n}^{(k_n)}(\nu,\nu_n)$ in boundary coefficients. It is defined by:
\begin{align}
& \int_{-\infty}^0 \nd \tau e^{ i k_0 \tau} (-\tau)^{\nu} h^{(1,2)}_{\nu_n}(a_n,-k_n\tau)\sim \text{C}_{\tilde{a}_n}^{(k_n)} x^{\nu+1} \, ,
\end{align}
while the left-hand side could be given by expanding the results of the 1-fold vertex integral family which we just obtained. 

For  $h^{(2)}_\nu(a,-k\tau)$ with real $\nu_n$, recall \eqref{eq:1V1fBCCk1inf}:
\begin{align}
&\text{C}_1^{(k_n)}(\nu,\nu_n) =\frac{\pi  2^{\nu-\nu _n+2} e^{i \pi \nu/2 } }
{ \left(1+e^{i \pi  \nu}\right)
\left(1+e^{i \pi  \left(\nu-2 \nu_n\right)}\right) 
\Gamma \left(\frac{1}{2}-\frac{\nu}{2}\right)
\Gamma\left(-\frac{\nu}{2}+\nu_n+\frac{1}{2}\right)}  \,,\nn\\
&\text{C}_2^{(k_n)}(\nu,\nu_n) =-\frac{i \pi  2^{\nu-\nu _n+2} e^{i \pi \nu /2} }
{\left(1-e^{i \pi  \nu}\right)
\left(1-e^{i \pi  \left(\nu-2 \nu_n\right)}\right) 
\Gamma \left(-\frac{\nu}{2}\right) 
\Gamma\left(-\frac{\nu}{2}+\nu_n+1\right)} \, . \label{eq:1Vkninfh2R}
\end{align}

For  $h^{(2)}_\nu(a,-k\tau)$ with imaginary $\nu$, we have
\begin{align}
&\text{C}_1^{(k_n)}(\nu,\nu_n) =\frac{\pi  2^{\nu -\nu _n+2} e^{\frac{1}{2} i \pi  \left(\nu -2 \nu _n\right)}}{\left(1+e^{i \pi  \nu }\right) \left(1+e^{i \pi ( \nu-2\nu_n) }\right) 
\Gamma\left(\frac{1}{2}-\frac{\nu }{2}\right)  
\Gamma \left(-\frac{\nu }{2}+\nu_n+\frac{1}{2}\right)}  \,,\nn\\
&\text{C}_2^{(k_n)}(\nu,\nu_n) =-\frac{ i \pi  2^{\nu -\nu _n+2} e^{\frac{1}{2} i \pi  \left(\nu -2\nu _n\right)}}
{\left(1-e^{i \pi  \nu }\right)
\left(1-e^{i \pi  \left(\nu -2 \nu _n\right)}\right) 
\Gamma \left(-\frac{\nu }{2}\right)
\Gamma \left(-\frac{\nu }{2}+\nu _n+1\right)} \, . \label{eq:1Vkninfh2I}
\end{align}

For  $h^{(1)}_\nu(a,-k\tau)$ with real or   imaginary $\nu$, the $\text{C}_{\tilde{a}_n}^{(k_n)}(\nu,\nu_n)$ should be the complex conjugate of \eqref{eq:1Vkninfh2R} when $\nu$ and $\nu_n$ are real:
\begin{align}
&\text{C}_1^{(k_n)}(\nu,\nu_n) =\frac{\pi  2^{\nu-\nu _n+2} e^{-i \pi \nu/2 } }
{ \left(1+e^{-i \pi  \nu}\right)
\left(1+e^{-i \pi  \left(\nu-2 \nu_n\right)}\right) 
\Gamma \left(\frac{1}{2}-\frac{\nu}{2}\right)
\Gamma\left(-\frac{\nu}{2}+\nu_n+\frac{1}{2}\right)}  \,,\nn\\
&\text{C}_2^{(k_n)}(\nu,\nu_n) =\frac{i \pi  2^{\nu-\nu _n+2} e^{-i \pi \nu /2} }
{\left(1-e^{-i \pi  \nu}\right)
\left(1-e^{-i \pi  \left(\nu-2 \nu_n\right)}\right) 
\Gamma \left(-\frac{\nu}{2}\right) 
\Gamma\left(-\frac{\nu}{2}+\nu_n+1\right)}    \, , \label{eq:1Vkninfh1}
\end{align}
and the result could be directly extended to the imaginary case.

With these results, we have completed the solving of the arbitrary $n$-fold vertex integral family of cosmological correlators.

\section{Properties of time-order $n$-vertex cosmological correlators}\label{sec:2-v}

While the two vertices linked by $G_{\pm\mp}$ are directly factorized as two integral, time-order propagators $G_{\pm\pm}$ combine two integrations of $\tau_i$ together and is likely to be much more complicated to solve. However, time-order propagators have elegant factorization properties, simplifying the IBP relation and differential equations \cite{Chen:2023iix}. This property further leads to the simplification of solutions as well. Thus, in this section, we will show the properties of time-order $n$-vertex cosmological correlators by
solving the integral family of tree-level 4-pt 2-vertex correlators as an example. The master integrals of this integral family for the $s$-channel with $G_{++}$ can be written as follows (with $k_{1;1}=k_{1;2}=k_s$ and $\nu_{1;1}=\nu_{1;2}=\nu_1$):
\ie
\text{I}_{\{a,b\}} \equiv \int \nd \tau_1 \nd \tau_2  (-\tau_1)^{\nu_{0}}e^{ik_{12}\tau_1}(-\tau_2)^{\nu_{0}}e^{ik_{34}\tau_2}h(\nu_1,a,-k_s\tau_1) \theta_{1,2}^{(1,1)} h(\nu_1,b,-k_s\tau_2)
\fe
and the remaining term
\ie
\text{I}_R=-\frac{4i}{\pi} e^{\pi \text{Im}\nu_1} (k_s)^{-2\nu_{1}-1} \int \nd \tau (-\tau)^{2\nu_0-2\nu_1}e^{i(k_{12}+k_{34})\tau}\, .
\fe
Here we have used the notation $k_{ij}=k_i+k_j$. In the following discussion, we will use $\bm{f}=\{f_i\}$ to denote the master integrals with $\text{I}_1= f_{\{0,0\}}$, $\text{I}_2= \text{I}_{\{0,1\}}$, $\text{I}_3= \text{I}_{\{1,0\}}$, $\text{I}_4= \text{I}_{\{1,1\}}$, and $\text{I}_5= \text{I}_R$. Notice that arbitrary cases of scalar or time derivative interaction are automatically included in the IBP system \cite{Chen:2023iix}. 

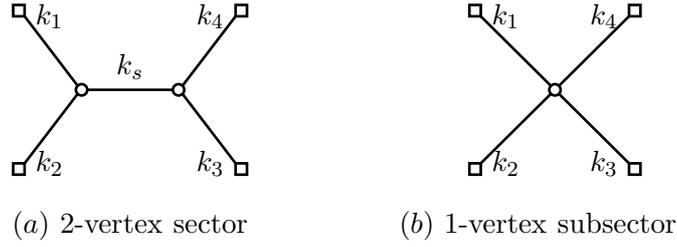
\begin{figure}

\centering
\begin{tikzpicture}[line width=1pt,scale=1.5]
\draw (0,0) circle (0.05);
\draw (0.85,0) circle (0.05);
\draw[fermionnoarrow] (-0.5,0.65)--(-0.03,0.03);
\draw[fermionnoarrow] (-0.5,-0.65)--(-0.03,-0.03);
\draw[fermionnoarrow] (0.05,0)--(0.8,0) ;
\draw[fermionnoarrow] (1.35,0.65)--(0.88,0.03);
\draw[fermionnoarrow] (1.35,-0.65)--(0.88,-0.03);
\draw (-0.6,0.75) rectangle (-0.5,0.65);
\draw (-0.6,-0.75) rectangle (-0.5,-0.65);
\draw (1.45,0.75) rectangle (1.35,0.65);
\draw (1.45,-0.75) rectangle (1.35,-0.65);
\node at (-0.5,0.65) [right] {$k_1$};
\node at (-0.5,-0.65) [right] {$k_2$};
\node at (1.35,0.65) [left] {$k_4$};
\node at (1.35,-0.65) [left] {$k_3$};
\node at (0.425,-1.2) {($a$) 2-vertex sector};
\node at (0.425,0) [above] {$k_s$};

\begin{scope}[shift={(4,0)}]
\draw (0.15,0) circle (0.05);
\draw[fermionnoarrow] (-0.5,0.65)--(0.12,0.03);
\draw[fermionnoarrow] (-0.5,-0.65)--(0.12,-0.03);
\draw[fermionnoarrow] (0.8,0.65)--(0.18,0.03);
\draw[fermionnoarrow] (0.8,-0.65)--(0.18,-0.03);
\draw (-0.6,0.75) rectangle (-0.5,0.65);
\draw (-0.6,-0.75) rectangle (-0.5,-0.65);
\draw (0.9,0.75) rectangle (0.8,0.65);
\draw (0.9,-0.75) rectangle (0.8,-0.65);
\node at (-0.5,0.65) [right] {$k_1$};
\node at (-0.5,-0.65) [right] {$k_2$};
\node at (0.8,0.65) [left] {$k_4$};
\node at (0.8,-0.65) [left] {$k_3$};
\node at (0,-1.2) {($b$) 1-vertex sub-sector};
\end{scope}

\end{tikzpicture}
\caption{ ($a$) is a diagram of the  2-vertex correlators, with the vertices could be scalar or time derivative interactions.  $\text{I}_{1,2,3,4}$ belong to this sector.  The IBP relation of integrals in ($a$) will automatically involve ($b$), which is given by pinching the propagator of ($a$) \cite{Chen:2023iix}. $\text{I}_5$ belong to this sub-sector. }
\end{figure}

\subsection{Differential equations}
The differential equations of the time-order 2-vertex integral family can be written as follows
\ie
\nd \textbf{I}=\nd \Omega \textbf{I}
\fe
with 5$\times$5 $\nd\log$ matrix
\ie
\Omega=\begin{pmatrix} \mathbf{A} & \mathbf{R}\\ \mathbf{0} & \mathbf{C}\end{pmatrix}.
\fe
and
\ie
&\mathbf{A}=\Omega_1(k_{12},k_s)\otimes \mathbf{1}_{2\times2}+\mathbf{1}_{2\times2}\otimes\Omega_1(k_{34},k_s),\,~~~~\nn\\
&\mathbf{C}=(-2 \nu_0+2\nu_1-1) \log (k_{12}+k_{34})+(-2 \nu_1-1) \log (k_s)\\
&\mathbf{R}=\left(
\begin{array}{c}
 \frac{1}{2} i (\log (k_{12}-k_{s})-\log (k_{12}+k_{s})+\log (k_{34}-k_{s})-\log (k_{34}+k_{s})) \\
 \frac{1}{2} (\log (k_{12}-k_{s})+\log (k_{12}+k_{s})-\log (k_{34}-k_{s})-\log (k_{34}+k_{s})) \\
 \frac{1}{2} (-\log (k_{12}-k_{s})-\log (k_{12}+k_{s})+\log (k_{34}-k_{s})+\log (k_{34}+k_{s})) \\
 \frac{1}{2} i (\log (k_{12}-k_{s})-\log (k_{12}+k_{s})+\log (k_{34}-k_{s})-\log (k_{34}+k_{s})) \\
\end{array}
\right)  \label{eq:dlogDE2-v}
\fe
where $\Omega_1(k_0,k_1)$ equals the matrix $\Omega$ defined in \eqref{eq:dlogDE1V1f}. Following the last section, we find that taking one of $k_i$ to $\infty$ of each vertex integral is likely to be a good choice for easily determining boundary conditions. However, note that there is a factor $1/(k_{12}+k_{34})$ in the differential equation. This factor leads to that $(k_{12},k_{34})=(\infty,\infty)$ a degenerate multivariate pole. To see that, consider $t_1=1/k_{12}, t_2=1/k_{34}$, then there are denominators $t_1$, $t_2$ and $t_1+t_2$ in the differential equations with respect to $t_1$ and $t_2$. The denominators equal to zero provide three hyper-surfaces across $(0,0)$ in this two-variable problem of $t_1$ and $t_2$. Thus, $(0,0)$ is a degenerate pole of $(t_1,t_2)$. By other words, $(\infty,\infty)$ is a degenerate pole of $(k_{12},k_{34})$. Hence, we need to apply blow-up to them when using the power series method around  $(k_{12},k_{34})=(\infty,\infty)$. Otherwise, the expansion will be illness and depend on expanding which variables first. As an alternative choice, we choose the transformation including blow-up to be $x=k_{34}/k_{12}$ and $y=1/k_{34}$. By this choice, $(x,y)=(0,0)$ is an non-degenerate pole. Now assuming all master integrals have a general form $$x^{\lambda}y^{\mu}\sum_{j,k=0}^{\infty}\text{C}(i,j,k)x^jy^k$$ with lowest weights $\lambda,\mu$ for $x$ and $y$, there are five non-trivial solution sets from the equation for C($i$,0,0):
\ie
&\{\text{C}(i\neq 1,0,0)= 0,\lambda= \nu_0+1,\mu= 2\nu_0+2\},\\
&\{\text{C}(i\neq 2,0,0)= 0,\lambda= \nu_0+1,\mu= 1+2\nu_0-2 \nu_1\},\\
&\{\text{C}(i\neq 3,0,0)= 0,\lambda= \nu_0-2 \nu_1,\mu= 1 + 2 \nu_0 - 2 \nu_1\},\\
&\{\text{C}(i\neq 4,0,0)= 0,\lambda= \nu_0-2 \nu_1,\mu= 2\nu_0-4 \nu_1\},\\
&\Big\{\text{C}(2,0,0)= \frac{\text{C}(5,0,0)}{2 \nu_1 -\nu_0},\text{C}(3,0,0)= \frac{\text{C}(5,0,0)}{\nu_0+1}, \\
&\text{C}(1,0,0)=\text{C}(4,0,0)= 0,\lambda=\mu= 2 \nu_0-2 \nu_1+1\Big\}
\fe
where the C($i$,0,0)s can be determined by boundary conditions. To avoid ambiguity, we will denote the non-zero coefficient in the $i$-th solution by C$^{[i]}$ in the following discussion such that
\ie
\bm{\text{I}}=\sum_{i=1}^{5}\text{C}^{[i]}\bm{f}^{[i]}
\fe
The first four general solutions are just direct products of solutions of the 1-fold general solutions. This is due to the structure of \textbf{A} in \eqref{eq:dlogDE2-v} as we will present more discussion in Sec \ref{sec:2-vfactoriztion}. To avoid confusion with symbols, let's re-denote the results of 1-vertex  by
\begin{align}
& V_{j}^{[i]}(x) = f_j^{[i]}(1/x,k_s) \, ,
\end{align}
where the $f_j^{[i]}$ are the ones given in \eqref{eq:1V1fsolseriesk0inf}.
We can write the first 4 solutions in a compact form:
\ie
f^{[\tilde{\bm{a}}]}_{\tilde{\bm{b}}}=V_{\tilde{b}_1}^{[\tilde{a}_1]}(xy)V_{\tilde{b}_2}^{[\tilde{a}_2]}(y)\, ,\ ~~f^{[\tilde{\bm{a}}]}_{5}= \bm{0} \, ,
\fe
with $\bm{a}=a_1,a_2$, $\bm{b}=b_1,b_2$, $a_i=0,1$, $b_i=0,1$.
For example, solution $\bm{f}^{[1]}$ is
\ie
&{ \bm{f}}^{[\{0,0\}]}=\left(\begin{array}{c}\left(\begin{array}{c}V_{\tilde{0}}^{[\tilde{0}]}(xy)\\V_{\tilde{1}}^{[\tilde{0}]}(xy)\end{array}\right)\otimes\left(\begin{array}{c}V_{\tilde{0}}^{[\tilde{0}]}(y)\\V_{\tilde{1}}^{[\tilde{0}]}(y)\end{array}\right)\\~~0\end{array}\right)=\left(\begin{array}{c}~~V_{\tilde{0}}^{[\tilde{0}]}(xy)~~V_{\tilde{0}}^{[\tilde{0}]}(y)~~\\
~~V_{\tilde{0}}^{[\tilde{0}]}(xy)~~V_{\tilde{1}}^{[\tilde{0}]}(y)~~\\
~~V_{\tilde{1}}^{[\tilde{0}]}(xy)~~V_{\tilde{0}}^{[\tilde{0}]}(y)~~\\
~~V_{\tilde{1}}^{[\tilde{0}]}(xy)~~V_{\tilde{1}}^{[\tilde{0}]}(y)~~\\
0\end{array}\right)\\
&=x^{\nu_0+1}y^{2\nu_0+2}\left(\begin{array}{c}\, _2\text{F}_1\left(\frac{\nu_0+1}{2},\frac{\nu_0+2}{2};\nu_1+1;k_s^2 y^2\right) \, _2\text{F}_1\left(\frac{\nu_0+1}{2},\frac{\nu_0+2}{2};\nu_1+1;k_s^2 x^2 y^2\right)\\
\frac{i k_s (\nu_0+1) y \, _2\text{F}_1\left(\frac{\nu_0+2}{2},\frac{\nu_0+3}{2};\nu_1+2;k_s^2 y^2\right) \, _2\text{F}_1\left(\frac{\nu_0+1}{2},\frac{\nu_0+2}{2};\nu_1+1;k_s^2 x^2 y^2\right)}{2 (\nu_1+1)}\\
\frac{i k_s (\nu_0+1) x y \, _2\text{F}_1\left(\frac{\nu_0+1}{2},\frac{\nu_0+2}{2};\nu_1+1;k_s^2 y^2\right) \, _2\text{F}_1\left(\frac{\nu_0+2}{2},\frac{\nu_0+3}{2};\nu_1+2;k_s^2 x^2 y^2\right)}{2 (\nu_1+1)}\\
\frac{- k_s^2 (\nu_0+1)^2x y^2  \, _2\text{F}_1\left(\frac{\nu_0+2}{2},\frac{\nu_0+3}{2};\nu_1+2;k_s^2 y^2\right) \, _2\text{F}_1\left(\frac{\nu_0+2}{2},\frac{\nu_0+3}{2};\nu_1+2;k_s^2 x^2 y^2\right)}{4 (\nu_1+1)^2}\\
0\end{array}\right) \label{eq:2-vhomosol}
\fe

The first 4 solutions have no contribution from the remaining integral $f_5$. Thus, they are homogeneous parts of solutions. The non-homogeneous solution related to non-zero $f_5$  also can be solved via power series expansion straightforwardly. It is given as follows:
\ie
&\bm{f}^{[5]}=x^{2\nu_0-2\nu_1+1}y^{2\nu_0-2\nu_1+1}\\
&~~~~~\times \left(\begin{array}{c}\sum_{m,n=0}^{\infty } \frac{-(-x)^m \left(\frac{1}{4} k_s^2 x y\right)^{n+1} (2 \nu_0-2 \nu_1+1)_{m+2 n+1} \left(4 i x^n y^n \right)}{4m! k_s  \left(\frac{\nu_0+1}{2}+\frac{m}{2}\right)_{n+1} \left(\frac{1}{2} (m+\nu_0-2 \nu_1+1)\right)_{n+1}}\\
\sum_{m,n=0}^{\infty } \frac{-(-x)^m  \left(\frac{1}{4} k_s^2 x^2 y^2\right)^n (2 \nu_0-2 \nu_1+1)_{m+2 n}}{2m!  \left(\frac{\nu_0+2}{2}+\frac{m}{2}\right)_n \left(\frac{1}{2} (m+\nu_0-2 \nu_1)\right)_{n+1}}\\
\sum_{m,n=0}^{\infty }\frac{(-x)^m (2 \nu_0-2 \nu_1+1)_{m+2 n}\left(\frac{1}{4} k_s^2 x^2 y^2\right)^n}{2m!  \left(\frac{\nu_0+1}{2}+\frac{m}{2}\right)_{n+1} \left(\frac{1}{2} (m+\nu_0-2 \nu_1+1)\right)_n}\\
\sum_{m,n=0}^{\infty } \frac{-(-x)^m \left(\frac{1}{4} k_s^2 x y\right)^{n+1} (2 \nu_0-2 \nu_1+1)_{m+2 n+1} \left(4 i x^n y^n \right)}{4m! k_s  \left(\frac{\nu_0+2}{2}+\frac{m}{2}\right)_{n+1} \left(\frac{1}{2} (m+\nu_0-2 \nu_1)\right)_{n+1}}\\
 \, (1+x)^{-1-2\nu_0+2\nu_1}\end{array}\right)\\
&=x^{2\nu_0-2\nu_1+1}y^{2\nu_0-2\nu_1+1}\\
&\times{\small\left(\begin{array}{c}\sum_{m=0}^{\infty } \frac{-ik_sxy(-x)^m   (2\nu_0-2\nu_1+1)_{m+1}}{m!   \left(\nu_0+m+1\right)\left(\nu_0-2 \nu_1+m+1\right)}{}_3\text{F}_2\left[ 
\left.\begin{matrix} 
\nu_0-\nu_1+\frac{m+2}{2},\nu_0-\nu_1+\frac{m+3}{2},1 \\
\frac{\nu_0+m+3}{2},\frac{\nu_0-2 \nu_1+m+3}{2} 
\end{matrix} \right| 
k_s^2x^2y^2 \right]\\
\sum_{m=0}^{\infty } \frac{-(-x)^m (2\nu_0-2\nu_1+1)_{m} }{m!   (\nu_0-2 \nu_1+m)}{}_3\text{F}_2\left[
\left.\begin{matrix}
\nu_0-\nu_1+\frac{m+1}{2},\nu_0-\nu_1+\frac{m+2}{2},1 \\ 
\frac{\nu_0+m+2}{2},\frac{\nu_0-2 \nu_1+m+2}{2}
\end{matrix} \right| 
k_s^2x^2y^2\right]\\
\sum_{m=0}^{\infty }\frac{(-x)^m (2\nu_0-2\nu_1+1)_{m} }{m!   (\nu_0+m+1)}{}_3\text{F}_2\left[
\left.\begin{matrix}
\nu_0-\nu_1+\frac{m+1}{2},\nu_0-\nu_1+\frac{m+2}{2},1\\ 
\frac{\nu_0+m+3}{2},\frac{\nu_0-2 \nu_1+m+1}{2}
\end{matrix} \right| 
k_s^2x^2y^2\right]\\
\sum_{m=0}^{\infty }\frac{-ik_sxy(-x)^m   (2\nu_0-2\nu_1+1)_{m+1}}{m!   \left(\nu_0+m+2\right)\left(\nu_0-2 \nu_1+m\right)}{}_3\text{F}_2\left[
\left.\begin{matrix}
\nu_0-\nu_1+\frac{m+2}{2},\nu_0-\nu_1+\frac{m+3}{2},1\\
\frac{\nu_0+m+4}{2},\frac{\nu_0-2 \nu_1+m+2}{2}
\end{matrix} \right| 
k_s^2x^2y^2\right]\\
 \, (1+x)^{-1-2\nu_0+2\nu_1}\end{array}\right)}
\fe
In the next subsection, we will show how to obtain the relative coefficients $\text{C}^{[i]}$.

\subsection{Boundary conditions}

When $x,y\to0$, only the lowest power terms dominate. One can expand the Hankel function around $t=0$ under this limit since the integrand will be localized at that point due to the exponential terms, and then integrate it to determine the relative coefficients C($i$,0,0) of these solution sets. We can start with the remaining term first since it can be integrated analytically. We will get
\ie
\text{C}^{[5]}=-\frac{4 i e^{\pi \text{Im}[\nu_1]} k_s^{-2 \nu_1-1} e^{i \pi  (\nu_1-\nu_0)} \Gamma (2 \nu_0-2 \nu_1+1)}{\pi }.
\fe
Notice that in solution $\bm{f}^{[5]}$, $f^{[5]}_2$ and  $f^{[5]}_3$ are also non-zero at the leading order. An alternative way to determine $\text{C}^{[5]}$ is expanding $\theta(\tau_i-\tau_j)$ in $\text{I}_2$ or  $\text{I}_3$ as $0-\tau_j\delta(\tau_i)$ or $1+\tau_i\delta(\tau_j)$ for the selected blow-up and computing the contribution of the $\delta$ function part.

For the other 4 coefficients, we need to expand the Hankel functions in integrands of $\textbf{I}$ to the leading order and then integrate. For example, the master integral I$_1$ has the following expression after Wick rotation:
\ie
\text{I}_1=-\int \nd \tau_1 \nd \tau_2  (i\tau_1)^{\nu_{0}}e^{\tau_1/(xy)}(i\tau_2)^{\nu_{0}}e^{\tau_2/y}h(\nu_1,0,ik_s\tau_1) \theta_{1,2}^{(1,2)} h(\nu_1,0,ik_s\tau_2)
\fe
Note that after taking $x,y\to0$ we must have $-1{ \ll}\tau_2<\tau_1<0$ due to the blow-up process. The theta function at the leading order will consequently be taken to 0 or 1. Using the expansion of the Hankel function \eqref{eq:expandha2R}, for real $\nu_1$ we have
\ie
&\text{C}^{[1]}
=-e^{-i \pi\nu_0}\Gamma (\nu_0+1)^2\text{C}_1^{*(k_0)}(\nu_1)\text{C}_1^{(k_0)}(\nu_1)\\
&\text{C}^{[2]}=
-ie^{-i \pi  (\nu_0-\nu_1)}  k_s^{-2 \nu_1-1} \Gamma (\nu_0+1)\Gamma (\nu_0-2 \nu_1)\text{C}_1^{*(k_0)}(\nu_1)\text{C}_2^{(k_0)}(\nu_1)\\
&\text{C}^{[3]}
=-ie^{-i \pi  (\nu_0-\nu_1)}  k_s^{-2 \nu_1-1} \Gamma (\nu_0+1)\Gamma (\nu_0-2 \nu_1)\text{C}_2^{*(k_0)}(\nu_1)\text{C}_1^{(k_0)}(\nu_1)\\
&\text{C}^{[4]}
=\left(k_s^2\right)^{-2 \nu_1-1} e^{-i \pi  (\nu_0-2 \nu_1)}\Gamma (\nu_0-2 \nu_1)^2\text{C}_2^{*(k_0)}(\nu_1)\text{C}_2^{(k_0)}(\nu_1).
\fe
Here, we use the $\text{C}_{\tilde{a}}^{(k_0)}$  in \eqref{eq:expandha2Iagain}, and the $\text{C}_{\tilde{a}}^{*(k_0)}(\nu_1)$ here is just the $\text{C}_{\tilde{a}}^{(k_0)}$  in \eqref{eq:expandha1}. Thus, one can find that these boundary coefficients are exactly the product of the corresponding 1-vertex ones.

\subsection{Factorization of homogeneous solutions}\label{sec:2-vfactoriztion}

For this first-order linear differential equation system with 5 master integrals, there must be 5 arbitrary constants (coefficients) that need to be fixed by boundary conditions. This means that there are always 5 independent solution sets. One can set $f_5=f_R=0$ and then the differential system will become 4 master integrals that satisfy the $\nd \log$ form differential equations \textbf{A} in \eqref{eq:dlogDE2-v}. By factorization of IBP and differential equations \cite{Chen:2023iix}, \textbf{A} consists of the 1-vertex 1-fold $\nd \log$ form differential equations. Obviously, they can be written as the product of the two general solutions of 1-vertex 1-fold, as we have shown in \eqref{eq:2-vhomosol}.  These solutions are also called ``homogeneous parts". In addition, the left one with $f_5\neq0$ will correspond to the ``non-homogeneous part". 

The coefficients for the homogeneous part can also be written as the product of coefficients of 1-vertex 1-fold solutions. When we consider the boundary conditions, by choosing blow-up, the theta function will be expanded as 0 or 1 at the leading order. For example, since there are $e^{i k_{12} \tau_1}$  and $e^{i k_{34} \tau_2}$ in the integrand, the limitation $k_{12}\gg 1$ and $k_{34}\gg 1$ together with Wick rotation lead to that only the region $|\tau_1|\ll 1$ and $|\tau_2|\ll 1$ could contribute. The blow-up further choose that $k_{12}\gg k_{34}$, thus only the region $|\tau_1|\ll |\tau_2|\ll 1$ could contribute. Obviously,  $\theta(\tau_1-\tau_2)=1$ in the region contribute. This will cause the integrals involving two times to be factorized and exactly equal to the product of 1-vertex 1-fold integrals at leading order: 
\ie
&\int \nd \tau_1\nd \tau_2(-\tau_1)^{\nu_{0}}e^{ik_{12}\tau_1}(-\tau_2)^{\nu_{0}}e^{ik_{34}\tau_2}h(\nu_1,a,-k_s\tau_1) \theta_{1,2}^{(1,2)} h(\nu_1,b,-k_s\tau_2)\\
\longrightarrow &\left[\int \nd \tau_1(-\tau_1)^{\nu_{0}}e^{ik_{12}\tau_1}h^{(1)}(\nu_1,a,-k_s\tau_1)\right]\left[\int \nd \tau_2(-\tau_2)^{\nu_{0}}e^{ik_{34}\tau_2} h^{(2)}(\nu_1,b,-k_s\tau_2)\right].
\fe
The coefficients determined by them will consequently be the product 1-vertex 1-fold boundary coefficients. 

Furthermore, this argument can also be generalized to general tree-level cases. One can set all or a part of sub-sectors to be zero first. Then one can choose a blow-up transformation. This leads to step functions becoming one or zero correspondingly. As a result, boundary coefficients determined by leading order expansion are factorized. These factorized general solutions together with factorized boundary coefficients give the factorized particular solutions that are exactly the product of several particular solutions of the vertices which come from dividing the diagram by cutting several propagators. The dividing is not necessary to be cutting all the propagators and it is not necessary to be  1-vertex at each part.  This is the factorization property of solutions of tree-level cosmological correlators. If one sets all integrals in sub-sectors to be zero, it corresponds to cutting all the propagators in the dividing. Then, the remaining solutions are homogeneous solutions at the top-sector level. They are the product of particular solutions of the vertex integral family we have given in \ref{sec:1-v_nf}.

\section{Summary and outlook}\label{sec:summary}

In this work, we use power series expansion to solve the $\nd \log$-form differential equations of cosmological correlators. It gives multivariate hypergeometric solutions. We also analyze the properties of integrands of the cosmological correlator and find the boundary conditions easy to solve for arbitrary vertex integral families. The solutions are given in \ref{sec:1-v_nf}. Analytic continuation of the series expansion solution is also discussed. We show the recommended numerical analytic continuation by differential equations is efficient, straightforward, and convenient.  We indicate that blow-up transformation could be applied to solve differential equations by the power series expansion around degenerate poles. With this technique, we also find the boundary conditions easy to determine for 2-vertex correlators, which is likely to work for arbitrary $n$-vertex correlators cases. Then, we give the particular solutions to this 2-vertex example. By this example, we also discuss the factorization property of homogeneous part solutions of $n$-vertex correlators, with which, one can easily give them as the product of the solutions of vertex integral family.

We want to remind the readers, that our results not only elucidate the mathematical structure of cosmological correlator, benefit their evaluation and related applications in cosmological phenomenology, but also offer new insights into computational techniques for evaluating integrals of perturbative QFT including flat cases, mainly in two aspects. Firstly, we indicate that using blow-up one could easily handle the degenerate pole of differential equations in multivariate limitation. Secondly, our results show a potential way toward analytic evaluation of field theory integrals beyond MPL: $\nd \log$-form differential equations. Let us call some background of analytic evaluation perturbative QFT here. \cite{Henn:2013pwa} carry out the canonical differential equations method, the most powerful analytic method currently. It is the IBP-based differential equation that takes a special form called canonical differential equations. which means the  differential equations is  proportional to dimension regulator $\ep$, or says $\ep$-form for short, and also a $\nd \log$-form at the same time. This method works for those integrals that are multiple polylogarithms (MPL) in the expansion of $\ep$. Even though, its scope includes all one-loop Feynman integrals and a large part of common multi-loop integrals in flat QFT. However, the developing phenomenology of particle physics still calls for analytic methods that can go beyond MPL. The most frequently considered method of generalizing canonical differential equations to these cases currently is keeping the differential equations in $\ep$-form  rather than $\nd \log$-form (see \cite{Pogel:2022vat,Bogner:2019lfa, Broedel:2018iwv, Adams:2018yfj} and their references for examples and discussions). In this case, the elliptic symbol or beyond appears in the elements of differential equations.  The emerging new functions and series solutions are studied. Unlike  $\nd \log$-form, coefficients in differential equations and master integrals are non-algebraic for these cases, thus they are more complicated. Finding such $\ep$-form usually is also more difficult than $\nd \log$-form cases. However, our paper results imply that keeping $\nd \log$-form  rather than $\ep$-form is another way worth to be considered. We find it easy to get all order series solutions to $\nd \log$-form differential equations in our example. This is partially due to the expansion of $\nd \log$-forms are simple, avoiding non-algebraic expressions. It is simple, especially for the case that the function in $\log$ is rational, as shown in \eqref{eq:1vexpansionDE}, since in such cases the log can always be regarded as $\nd \log(z-c)$ for a selected
 parameter $z$.

 Our work leads to many topics that could be explored in the future. We merely list a part of them. Firstly, one can apply our methods to the correlators important to phenomenology that have not been evaluated. Even for the loop level, the framework of IBP already has been discussed in \cite{Chen:2023iix}. Although $\nd \log$-form differential equations may not be easy to get in this case,  we want to remind people, the (generalized) power series expansion method does not rely on $\nd \log$-form differential equations. 
The efficiency of this method for differential equations in general (usually rational) form has been confirmed in \cite{Moriello:2019yhu} at the beginning, and a subsequent series of works following it have also validated this. 
Our Sec \ref{sec:1V1fcontinue} confirm this as well. Automatic tools of this method, which have been widely applied in flat QFT, could also help,  for example, DiffEXP \cite{Hidding:2020ytt}.

Secondly, although in this work, we show all order expressions of multivariate hypergeometric solutions, the attainment of this result is partially dependent on the special structure of the dlog differential equations we have computed. For general $\nd \log$-form differential equations, could we find a formula to determine its all-order power series solution or express it as multivariate hypergeometric functions like in this paper? Alternatively, one can also explore that is there exists an algorithm, that is more efficient than the naive (generalized) power series expansion, to evaluate master integrals in the special case of $\nd \log$-form differential equations. Then, since one can quickly get numerical results at any point of phase space and analyze asymptotic behavior around the arbitrary singularity, this algorithm together with $\nd \log$-form matrix defines a series of new ``analytic functions".

\section*{Acknowledgements}
This work is supported by Chinese NSF funding under Grant No.11935013, No.11947301, No.12047502 (Peng Huanwu Center), No.12247103, No.U2230402, and China Postdoctoral Science Foundation No.2022M720386. YT is partly supported by the National Key R\&D Program of China (NO. 2020YFA0713000).







\bibliographystyle{JHEP}
\bibliography{references}

\end{document}